\documentstyle[11pt]{article}
\begin{document}

\hfill
\renewcommand{\thefootnote}{\fnsymbol{footnote}}
NIKHEF 96-010\footnote{The complete postscript file of this preprint
is available via anonymous ftp at nikhefh.nikhef.nl as
/pub/preprints/96-010.ps.Z }

\addtocounter{footnote}{-1}
\renewcommand{\thefootnote}{\arabic{footnote}}
\hfill
April 1996

\begin{center}
{\huge \bf The 3-loop QCD calculation of the
moments of deep inelastic structure functions  }\\ [8mm] S.A.
Larin$^{ab}$, P. Nogueira$^{c}$\footnote{
 Partially supported by Junta Nacional de Investiga\c{c}\~{a}o
 Cient\'{\i}fica e Tecnol\'{o}gica, Lisbon } , 
  T. van Ritbergen$^a$, J.A.M. Vermaseren$^a$ \\ [3mm]
\begin{itemize}
\item[$^a$]
 NIKHEF-H, P.O. Box 41882, \\ 1009 DB, Amsterdam, The Netherlands \\
\item[$^b$]
 Institute for Nuclear Research of the
 Russian Academy of Sciences   \\
 60th October Anniversary Prospect 7a,
 Moscow 117312, Russia
\item[$^c$]
 Instituto Superior T\'{e}cnico, \\
 Edif\'{\i}cio Ci\^{e}ncia ( F\'{\i}sica ) \\
 P-1096 Lisboa Codex, Portugal \\
\end{itemize}
\end{center}

\begin{abstract}
We present the analytic next-to-next-to-leading perturbative QCD corrections
in the leading twist approximation for the moments $N=2,4,6,8$ of the
flavour singlet deep inelastic structure functions $F_2$ and $F_L$.
We calculate the three-loop anomalous dimensions of the corresponding
singlet operators and the three-loop coefficient functions of
the structure functions $F_L$ and $F_2$. In addition, we obtained the
10$^{th}$ moment for the non-singlet structure functions in the same
order of perturbative QCD. We perform an analysis of the obtained results.
\end{abstract}

\section{Introduction}

The calculation of the next-to-next-to-leading (NNL) QCD approximation for
the structure functions $F_2$ and $F_L$ of deep inelastic electron-nucleon
scattering is important for the understanding of perturbative QCD and
for an accurate comparison of perturbative QCD with experiment. 
To obtain the NNL approximation for these structure functions in the
operator product expansion (OPE) formalism 
one needs the 3-loop anomalous dimensions of the operators, 
the 2-loop Wilson coefficient functions for $F_2$ 
and the 3-loop coefficient functions for $F_L$.
At present, these structure functions are known in the next-to-leading
approximation only, since the 3-loop anomalous dimensions and the 3-loop
coefficient functions for $F_L$ were not calculated yet.

The 1-loop anomalous dimensions were calculated in Ref. \cite{grwilcz}.
The complete 1-loop  coefficient functions  were obtained in Ref. \cite{bbdm}
(see also the references therein).
Anomalous dimensions in 2-loop order were obtained in 
Refs. \cite{frs}-\cite{hn} and the 2-loop coefficient functions 
were calculated in Refs. \cite{ddks}-\cite{lvm}. 
In a previous paper \cite{weNS} we presented the
 NNL corrections of the {\em non-singlet} type 
in the leading twist approximation for the moments $N=2,4,6,8$ of the
 deep inelastic structure functions $F_2$ and $F_L$.

In the present paper we calculate the NNL QCD corrections to the {\em singlet}
moments $N=2,4,6,8$ of both structure functions $F_2$ and $F_L$.
To this end, we calculate the corresponding 3-loop anomalous dimensions
and the 3-loop coefficient functions for the structure function $F_L$.
In addition, we present the 3-loop coefficient functions for the structure 
function $F_2$ for $N=2,4,6,8$.
We also obtain the N=10 {\em non-singlet} moments of $F_2$ and $F_L$. 
The calculations are done for the leading twist approximation 
for zero quark masses.

In the fifth part of this paper we analize the effects that the calculated 
3-loop corrections have on the structure functions.

\section{Formalism}

We need to calculate the hadronic part of the amplitude for unpolarized 
deep inelastic electron-nucleon scattering which is given by the hadronic 
tensor
\begin{eqnarray}
\label{htensor}
W_{\mu \nu}(p,q) & = & \frac{1}{4\pi}
     \int d^4ze^{iq\cdot z}\langle p, {\rm nucl } \vert
 J_{\mu}(z)J_{\nu}(0)\vert\ {\rm nucl}, p\rangle
 \nonumber \\ & = &
 e_{\mu\nu}\frac{1}{2x}F_L(x,Q^2)+d_{\mu\nu}\frac{1}{2x}F_2(x,Q^2)
 \\
 e_{\mu\nu} & = & \left(g_{\mu \nu}-\frac{q_{\mu} q_{\nu}}{q^2}\right)
   \nonumber \\
 d_{\mu\nu} & = & \left(-g_{\mu \nu}-p_{\mu}p_{\nu}\frac{4x^2}{q^2}
              -(p_{\mu}q_{\nu}+p_{\nu}q_{\mu})\frac{2x}{q^2}\right) ,
 \nonumber
\end{eqnarray}
where $J_{\mu}$ is the electromagnetic quark current,
$x=Q^2/ (2p\cdot q)$ is the Bjorken scaling variable ($0 < x \leq 1$),
 $Q^2=-q^2$ is the transferred momentum and $\vert\ {\rm nucl}, p\rangle$ is
the nucleon state with momentum $p$. Spin averaging is assumed.
The longitudinal structure function $F_L$ is related to the structure function
 $F_1$ by $ F_L = F_2 -2xF_1$.

As one approaches the Bjorken limit, $Q^2 \rightarrow \infty$,
$ x$ fixed, one can show that the integration region in Eq. (\ref{htensor}) 
near the light cone $z^2\approx 0$ progressively dominates\cite{ioffe},
 due to increasingly rapid phase fluctuations 
of the term $e^{iq\cdot z}$ outside the light cone region
(and presuming that the integrand 
 $\langle p, {\rm nucl} \vert J_{\mu}(z)J_{\nu}(0)\vert\ 
 {\rm nucl }, p\rangle$
varies smoothly outside the light cone). Since we have to deal with
this non-local limit $z^2\approx 0$, a formal operator product expansion 
in terms of local operators can only be applied together with the
dispersion relation technique \cite{christ}.
These techniques together provide a systematic way to
study\footnote{For reviews see Refs. \cite{buras,reya,collins}.} 
the leading and non-leading contributions to the hadronic tensor.

The tensor $W_{\mu \nu}$ is, by application of the optical theorem,
related to a scattering amplitude $T_{\mu \nu}$ which is a more convenient
quantity for practical calculations since it has a time ordered
product of currents to which standard perturbation theory applies
($T_{\mu \nu}$ is the amplitude for forward elastic photon--nucleon 
 scattering )
\begin{equation} \label{optical}
 W_{\mu \nu}(p,q) = \frac{1}{2\pi} \mbox{Im}\, T_{\mu \nu}(p,q),\hspace{1cm}
 T_{\mu \nu}(p,q) = i \int d^4ze^{iqz}
\langle p, {\rm nucl } \vert
 T \left( J_{\mu}(z)J_{\nu}(0) \right) \vert\ {\rm nucl}, p\rangle .
\end{equation}
The operator product expansion in terms of local 
operators for a time ordered product of 
the two electromagnetic hadronic currents reads
\[
 i\int d^4ze^{iqz}T\left( J_{\nu_1}(z)J_{\nu_2}(0)\right) 
         = \hspace{10cm} \]
\[
 \sum_{N,j} (\frac{1}{Q^2})^N 
\left[\left(g_{\nu_1\nu_2}-\frac{q_{\nu_1}q_{\nu_2}}{q^2}\right)
 q_{\mu_1}q_{\mu_2}
 C_{L,N}^{j}(\frac{Q^2}{\mu^2},a_s) \right. 
-\left(g_{\nu_1 \mu_1}g_{\nu_2\mu_2}q^2  -g_{\nu_1\mu_1}q_{\nu_2}q_{\mu_2}
 \right.  \] 
\[ \left. \left.
   -g_{\nu_2\mu_2}q_{\nu_1}q_{\mu_1}   +g_{\nu_1\nu_2}q_{\mu_1}q_{\mu_2} \right)
      C_{2,N}^{j}(\frac{Q^2}{\mu^2},a_s)\right] 
q_{\mu_3}...q_{\mu_N} O^{j,\{\mu_1,...,\mu_N\}}(0)  
 \mbox{+ higher twists, } \]
\begin{equation} \hspace{13cm} \mbox{ j=$\alpha,\psi$,G} \label{OPE}
\end{equation}
where everything is assumed to be renormalized (with $\mu$ being
the renormalization scale).
    The use of the OPE in the short distance regime ( $z\rightarrow 0$ )
    differs from its use in the light cone region.
    In the former case the sum over spin-N extends to a finite value
    for a given approximation, while in the latter (the one we have to
    deal with) the sum over N extends to infinity.
The sum over N runs over the standard set of the spin-N twist-2 irreducible
(i.e. symmetrical and traceless in the indices $\mu_1,\cdots,\mu_N$)
flavour non-singlet quark operators and the singlet quark and gluon
operators:
\begin{eqnarray} \label{defoperatorns}
O^{\alpha,\{\mu_1,\cdots ,\mu_N\}} & = & \overline{\psi}\lambda^{\alpha}
  \gamma^{\{\mu_1}D^{\mu_2}\cdots D^{\mu_N\}}\psi,~~\alpha=1,2,...,(n_f^2-1) \\
O^{\psi,\{\mu_1,\cdots ,\mu_N\}} & = & \overline{\psi}
  \gamma^{\{\mu_1}D^{\mu_2}\cdots D^{\mu_N\}}\psi, \label{defoperatorquark}\\
O^{G,\{\mu_1,\cdots ,\mu_N\}} & = & G^{\{\mu\mu_1} D^{\mu_2}\cdots
  D^{\mu_{N-1}} G^{\mu_N \mu \} \label{defoperatorgluon} }.
\end{eqnarray}
   Here and in the following we denote the generators of
    the flavour group $SU(n_f)$ by $\lambda^{\alpha}$, and
    the covariant derivative by $D^{\mu_i}$; in addition,
    it is understood that the symmetrical and traceless part
    is taken with respect to the indices in curly brackets.
    The functions $C_{k,N}^{j}(Q^2/\mu^2,a_s)$
    are the coefficient functions for the above operators.
Since the coefficient functions $C_{k,N}^{\alpha}$ of non-singlet
operators depend trivially on the number $\alpha$ (see e.g. Ref. 
\cite{weNS} or section 5 of the present article)  we will use for them the 
standard notation  $C_{k,N}^{ns}$.
Here and throughout the whole paper we use the notation
\begin{equation} \label{alphadef}
 a_s=\frac{g^2}{16\pi^2}=\frac{\alpha_s}{4\pi} 
\end{equation}
 for the QCD strong coupling constant.
The direct application of the OPE of Eq. (\ref{OPE})
 to the Green function $T_{\mu\nu}$
leads to  a formal expansion for $T_{\mu\nu}$ in terms of the variable
$ q\cdot p/Q^2 = 1/(2x)$  i.e. an expansion for unphysical 
$x \rightarrow \infty$, 
\[
i \int d^4ze^{iqz}
 \langle p, {\rm nucl } \vert
 T \left( J_{\mu}(z)J_{\nu}(0) \right) \vert\ {\rm nucl}, p \rangle 
         = \hspace{8cm} \]
\[
 \sum_{N,j} (\frac{1}{2 x})^N
\left[\left( g_{\mu \nu}-\frac{q_{\mu} q_{\nu}}{q^2}  \right)
 C_{L,N}^{j}\left(\frac{Q^2}{\mu^2},a_s\right) \right.
-\left(
g_{\mu \nu}+p_{\mu}p_{\nu}\frac{4x^2}{q^2}
 \right.  \] 
\newpage
\[ \left. \left. +(p_{\mu}q_{\nu}+p_{\nu}q_{\mu})\frac{2x}{q^2}
    \right)
      C_{2,N}^{j}\left(\frac{Q^2}{\mu^2},a_s\right)\right]
  A_{{\rm nucl},N}^j(m_n^2/\mu^2)
 \mbox{+ higher twists, } \]
\begin{equation} \hspace{13cm} \mbox{ j=$ns,\psi$,G}
 \label{overx} \end{equation}
where the spin averaged matrix elements are defined as
\begin{equation}
\label{mel}
\langle p , {\rm nucl} \vert O^{j, \{\mu_1,...,\mu_N\}}\vert
 {\rm nucl} , p\rangle
=p^{\{\mu_1}...p^{\mu_N\}}A_{{\rm nucl},N}^j(m_n^2/\mu^2)
\end{equation}
and $m_n$ is the nucleon mass.

To perform the proper analytic continuation of the representation 
Eq. (\ref{overx}) to the  physical region
$0 < x \leq 1$ one applies a dispersion relation in the complex $x$ plane
to the Green function $T_{\mu\nu}$. For electron-nucleon scattering 
where we have hermitian currents $J^{\mu}$ one finds that the Mellin moments 
of the structure functions $F_k$ are expressed through the parameters of the 
operator product expansion (\ref{OPE})
\begin{equation} \label{Mkn}
\frac{1+(-1)^N}{2} M_{k,N} \equiv 
\frac{1+(-1)^N}{2} \int_0^1 dx\hspace{1mm} x^{N-2} F_k(x,Q^2) 
= \sum_{i=ns,\psi,G}
  C^i_{k,N} \left(\frac{Q^2}{\mu^2},a_s\right) 
A_{{\rm nucl},N}^i   
\end{equation}
Please note that the odd Mellin moments of $F_k$ are not fixed by this equation.
However, all moments in the complex N plane are fixed by analytic
continuation from the even Mellin moments when all the even moments are known.
This means that the structure functions in $x$-space, $0 < x \leq 1$,
 can be found by means of the inverse Mellin transformation when the
 (infinite set of) even moments are known.

The $Q^2$-dependence of the coefficient functions  can be studied by the
use of the renormalization group equations
\begin{eqnarray} 
\left[ \mu^2 \frac{\partial}{\partial \mu^2} + \beta(a_s(\mu^2)) 
 \frac{\partial}{\partial a_s(\mu^2)} \right] 
C^i_{k,N} \left(\frac{Q^2}{\mu^2},a_s(\mu^2) \right)
&= &  C^j_{k,N} \left(\frac{Q^2}{\mu^2},a_s(\mu^2) \right) 
   \gamma_N^{ji}(a_s(\mu^2)) \nonumber \\
 \label{callansinglet} 
& &\hspace{3.5cm} , \mbox{ i,j=$\psi$,G}  \\ \label{callannonsin}
\left[ \mu^2 \frac{\partial}{\partial \mu^2} + \beta(a_s(\mu^2)) 
 \frac{\partial}{\partial a_s(\mu^2)} \right]
C^{ns}_{k,N} \left(\frac{Q^2}{\mu^2},a_s(\mu^2) \right)
 &= &  C^{ns}_{k,N} \left(\frac{Q^2}{\mu^2},a_s(\mu^2) \right)
 \gamma_N^{ns} (a_s(\mu^2))
\end{eqnarray}
where Eq. (\ref{callansinglet}) represents the 
singlet sector where quark and gluon operators
 mix under renormalization, and Eq. (\ref{callannonsin})
 is the non-singlet equation.
$\beta (a)$ is the beta-function that determines the renormalization scale
dependence of the renormalized coupling constant. It is known at 
three loops \cite{tvz} in the $\overline{\rm MS}$ scheme
\begin{eqnarray}
 \frac{ \partial a_s}{\partial \ln \mu^2} &\equiv &\beta(a_s)  =
  - \beta_{0} a_s^2 - \beta_{1} a_s^3 - \beta_{2} a_s^4 + O(a_s^5), \nonumber \\
\beta_{0} & = & ( \frac{11}{3} C_A - \frac{4}{3} T_F n_f )
 \nonumber \\
 \beta_{1} & = &
( \frac{34}{3}C_A^2 - 4 C_F T_F n_f -\frac{20}{3} C_A T_F n_f
 )  \nonumber \\
 \beta_{2} & = & ( \frac{2857}{54} C_A^3
 +2 C_F^2 T_F n_f - \frac{205}{9} C_F C_A T_F n_f - \nonumber \\
 & & - \frac{1415}{27} C_A^2 T_F n_f
 + \frac{44}{9} C_F T_F^2 n_f^2
  + \frac{158}{27} C_A T_F^2 n_f^2 ) \label{betafunction}
\end{eqnarray} 
where $C_{F} = \frac{4}{3}$ and $C_{A}= 3$ are the Casimir operators
of the fundamental and
adjoint representations of the colour group $SU(3)$,
$T_{F} = \frac{1}{2}$ is the
trace normalization of the fundamental representation and $n_f$ is the number
of (active) quark flavours.
The anomalous dimensions $ \gamma_N (a_s)$  
determine the renormalization scale dependence of the operators, that is
\begin{eqnarray}\label{defgammas}
\frac{d }{d \ln \mu^2 } O_{R}^{j, \{\mu_1,...,\mu_N\}}
 & \equiv & -  \gamma_N^{ji} (a_s) \hspace{.1cm} 
 O_{R}^{i, \{\mu_1,...,\mu_N\}}
 ,\hspace{1cm} \mbox{ i,j=$\psi$,G} \\ \label{defgammans}
 \frac{d }{d \ln \mu^2 } O_{R}^{ns, \{\mu_1,...,\mu_N\}}
 & \equiv & - \gamma_N^{ns} (a_s) \hspace{1mm}
 O_{R}^{ns, \{\mu_1,...,\mu_N\}}.
\end{eqnarray} 
We define renormalized operators in terms of bare operators
as $ O_{R} = Z\, O_{B} $ and find
\begin{equation} \label{defren}
 \frac{d }{d \ln \mu^2 } O_{R}
 = \left( \frac{d }{d \ln \mu^2 } Z\right) O_{B}
 =  \left( \frac{d }{d \ln \mu^2 } Z\right) Z^{-1} O_{R} 
 \hspace{.5cm} \Rightarrow\hspace{.5cm}
 \gamma = - \left( \frac{d }{d \ln \mu^2 }
    Z\right) Z^{-1} 
\end{equation}
where it is understood that in the singlet case $Z$ represents a
matrix $Z^{ij}$.
The renormalization group equations are solved in the standard form
\begin{eqnarray}
\label{rg}
C_{k,N}^{ns} \left(\frac{Q^2}{\mu^2},a_s(\mu^2)\right)
& = & C_{k,N}^{ns} (1,a_s(Q^2))
\times \exp\left(-\int_{a_s(\mu^2)}^{a_s(Q^2)}da'_s\,
\frac{\gamma_N^{ns} (a'_s)}{\beta(a'_s)}\right)  
\end{eqnarray}
The solution for the singlet equations has a similar form but 
since one gets the exponential of a matrix of anomalous 
dimensions one has to define the exponential properly 
in the singlet case
(i.e. a T-ordered exponential, see e.g. Ref. \cite{veltman}).
Here $a_s(Q^2) \equiv a_s(Q^2/\Lambda_{\overline{\rm MS}}^2)$ is 
the renormalized 
(i.e. running) coupling constant at the renormalization scale $Q^2$ 
\begin{eqnarray}
\label{alphaeff}
        a_s \left( \frac{Q^2}{\Lambda_{\overline{\rm MS}}^2}\right) & = &
        \frac{1}{\beta_0 \ln\left(\frac{Q^2}{\Lambda_{\overline{\rm MS}}^2}
 \right)}
        - \frac{\beta_1}{\beta_0^3}
        \frac{\ln\  \ln\left(\frac{Q^2}{\Lambda_{\overline{\rm MS}}^2}
 \right)}{
       \ln^2\left(\frac{Q^2}{\Lambda_{\overline{\rm MS}}^2}\right)}
  \nonumber \\
     && + \frac{1}{\beta_0^5 \ln^3\left(\frac{Q^2}{\Lambda_{\overline{\rm
  MS}}^2}
          \right)}
        \left( \beta_1^2 \ln^2\ln\left(\frac{Q^2}{\Lambda_{\overline{\rm
  MS}}^2}
  \right)
        - \beta_1^2 \ln \ \ln\left(\frac{Q^2}{\Lambda_{\overline{\rm 
 MS}}^2}\right)
        + \beta_2\beta_0 - \beta_1^2 \right)
\end{eqnarray}
and $\Lambda_{\overline{\rm MS}}$ is the fundamental 
 scale of QCD in the
$\overline{\rm MS}$-scheme.
In practice one may use the DGLAP evolution equations \cite{glap} for 
matrix elements of operators at the scale $\mu^2 = Q^2$ 
(i.e. $Q^2$-dependent parton distributions) 
instead of the renormalization group 
equations for the coefficient functions 
(for perturbative solutions of the DGLAP equations in moment space
see e.g. Refs. \cite{frs,kellis}).

\section{The method}

In this section we will discuss the method \cite{glt} for the calculation of 
anomalous dimensions and coefficient functions in considerable detail
 as it applies to the singlet sector. Let us first elaborate on some details
specific to the dimensional regularization\cite{hv} and the minimal 
subtraction scheme \cite{h}, and it's standard modification, the 
$\overline{\rm MS}$-scheme \cite{bbdm}, which form a modern basis for multiloop 
calculations in QCD.
We use the symbol $a_s$ for the renormalized coupling 
constant and $a_b$ for the bare coupling constant.   
Although renormalization constants Z contain poles in $\varepsilon$ in
 D$=4-2\varepsilon$ dimensions,
anomalous dimensions are finite as D$\rightarrow$ 4. 
This fact gives expressions for the higher poles of Z in terms of the first 
poles of Z.
To see this we write Eq. (\ref{defren}) as
\begin{equation} \label{zcondition}
 \gamma  Z = - \left( \frac{d }{d \ln \mu^2 }
    Z(a_s,\frac{1}{\varepsilon}) \right) =
 - \left( \frac{\partial }{\partial a_s} Z \right) 
  \frac{d a_s}{d \ln \mu^2 } 
 =  - \left( \frac{\partial }{\partial a_s} Z \right) \left [
 -\varepsilon a_s + \beta (a_s) \right] 
\end{equation}
where  $\beta (a_s)$ is the 4-dimensional beta function 
of Eq. (\ref{betafunction}) and $[ -\varepsilon a + \beta (a)]$ is the
 beta function in $4-2\varepsilon$ dimensions. This latter function
 receives no higher order corrections in $\varepsilon$ due to
the form of renormalization factors in the minimal subtraction scheme, viz.: 
\[ a_{b} = Z_{a_s} a_s  \, ,
   \hspace{1cm} Z_{a_s} = 
        1 - \frac{\beta_0}{\varepsilon} a_s
          + (\frac{\beta_0^2}{\varepsilon^2}
         - \frac{\beta_1}{2 \varepsilon}) a_s^2+ O(a_s^3) \]

\[ \frac{ d ( a_{b} \mu^{2\varepsilon})}{d \ln\mu^2} = 0 
   = \varepsilon Z_{a_s} a_s \mu^{2\varepsilon} +
        \frac{\partial Z_{a_s} }{\partial a }
         \frac{d a_s}{d \ln \mu^2 } a_s \mu^{2\varepsilon}
        + Z_{a_s} \frac{d a_s}{d \ln \mu^2 } \mu^{2\varepsilon} \]
\begin{equation} \Rightarrow
 \frac{d a_s}{d \ln \mu^2 } = - \frac{ \varepsilon Z_{a_s} a_s}{
      \frac{\partial Z_{a_s} }{\partial a_s } a_s + Z_{a_s} }  =
  -\varepsilon a_s + \beta (a_s) \end{equation}
where $a_b$ is a dimensionless object and $a_{b} \mu^{2\varepsilon}$ is
the bare coupling constant which is invariant under the renormalization group
transformations.
The factors $Z^{ij}$ are calculated as series in $a_s$, 
and have the well known form
\[ Z^{ij} = Z^{ij(0)} + Z^{ij(1)}(a_s)/ \varepsilon
                           + Z^{ij(2)}(a_s)/ \varepsilon^2  + \cdots 
,\hspace{1cm} \mbox{$i,j = \psi,G$} \]
 with $Z^{\psi\psi(0)}=Z^{GG(0)}=1$, $Z^{\psi G(0)}=Z^{G\psi(0)}=0$. 
It is helpful to write the matrix equation (\ref{zcondition})
as 4 separate equations,
\begin{eqnarray}
 \underline{ \gamma^{\psi\psi} Z^{\psi\psi} } + \gamma^{\psi G} Z^{G\psi} & = &
 - \left[\hspace{1mm} \underline{ -\varepsilon a_s} + \beta (a_s) \right]
  \frac{\partial }{\partial a_s} \underline{ Z^{\psi\psi} } \nonumber \\
  \gamma^{\psi\psi} Z^{\psi G} + \underline{ \gamma^{\psi G} Z^{GG} } & = &
 - \left[\hspace{1mm} \underline{ -\varepsilon a_s} + \beta (a_s) \right]
  \frac{\partial }{\partial a_s} \underline{ Z^{\psi G}} \nonumber \\
 \underline{ \gamma^{G\psi} Z^{\psi\psi}} + \gamma^{GG} Z^{G\psi} & = &
 - \left[ \hspace{1mm} \underline{ -\varepsilon a_s} + \beta (a_s) \right] 
  \frac{\partial }{\partial a_s} \underline{ Z^{G\psi}  } \nonumber \\
 \gamma^{G\psi} Z^{\psi G} + \underline{ \gamma^{GG} Z^{GG} } & = &
 - \left[ \hspace{1mm} \underline{ -\varepsilon a_s } + \beta (a_s) \right]
  \frac{\partial }{\partial a_s} \underline{ Z^{GG} } \label{spelledout}
\end{eqnarray}
where we underlined the terms that contribute to the
lowest order in $\varepsilon$ (i.e. order $\varepsilon^0$). From these terms
one immediately finds that the anomalous dimensions are expressed through
the coefficients in front of the first poles of $Z$
\begin{equation}
 \gamma^{ij} =  a_s \left( 
       \frac{\partial }{\partial a_s} Z^{ij(1)}(a_s) \right) 
 \hspace{1cm} \mbox{$i,j = \psi,G$} 
\end{equation}
where $Z^{ij(1)}(a_s)$ was defined as the order
 $1/ \varepsilon $ part of $Z^{ij}$.
The coefficients of higher poles in $Z$ can then be expressed in terms of 
$\gamma^{ij}$ by substituting the expression for $\gamma^{ij}$ back into
equation (\ref{spelledout}).

The operator product expansion of Eq. (\ref{OPE}) is an operator statement
and both the coefficient functions $C_{k,N}^i$ and the anomalous dimensions
$\gamma_N^{ij}$ of the operators are functions and do therefore not depend
on the hadronic states of the Green function to which one wishes to apply
the OPE. The information on the hadronic target is contained in the 
operator matrix elements $A_{N}^i$ in Eq. (\ref{mel}) 
which are generally not calculable perturbatively.
It is therefore standard to consider simpler Green functions with
quarks and gluons as external particles, instead of the physical
nucleon states, in the calculation of 
 coefficient functions and anomalous dimensions.
In this case the Green functions can be calculated in perturbation 
theory as well as the operator matrix elements
and the anomalous dimensions 
and coefficient functions can be extracted
as will be shown below in detail. 

Let us consider the following 4-point Green functions
\begin{equation}
 T_{\mu \nu}^{\rm q \gamma q \gamma}(p,q) = i \int d^4ze^{iqz}
\langle p, {\rm quark } \vert
 T \left( J_{\mu}(z)J_{\nu}(0) \right) \vert\ {\rm quark}, p\rangle 
\end{equation}
\begin{equation}
 T_{\mu \nu}^{\rm g \gamma g \gamma}(p,q) = i \int d^4ze^{iqz}
\langle p, {\rm gluon } \vert
 T \left( J_{\mu}(z)J_{\nu}(0) \right) \vert\ {\rm gluon}, p\rangle 
\end{equation}
where the label $\gamma$ is used to indicate an external photon,
q indicates an external quark and g an external gluon. 
Spin and colour averaging  for the quark and gluon  states is assumed.
Analogously to the decomposition of the hadronic tensor $W_{\mu\nu}$ in terms
of $F_2$ and $F_L$ we decompose the Green functions $T_{\mu \nu}$ in terms of 
$T_L$ and $T_2$. In the leading twist approximation (i.e. dropping
non-leading terms in $p^2$) one finds
\begin{equation}
 T_L = \frac{-q^2}{(p\cdot q)^2}p^\mu p^\nu \hspace{1mm}T_{\mu\nu} ,\hspace{1cm} 
   T_2 = -
 \left( \frac{3-2\varepsilon}{2-2\varepsilon}\hspace{1mm} 
  \frac{ q^2}{(p\cdot q)^2} p^\mu p^\nu
   + \frac{1}{2-2\varepsilon} \hspace{1mm} g^{\mu\nu} \right) T_{\mu\nu}. 
\end{equation}
Applying the OPE to $T^{\rm q \gamma q \gamma}$
 and $T^{\rm g \gamma g \gamma}$ we find the
following  equations for the renormalized Green functions 
\[
 T^{\rm q \gamma q \gamma}_{k}(p,q,a_s,\mu^2,\varepsilon) = 
   \hspace{13cm} \]
\[   \sum_{N=2}^{\infty} \left( \frac{1}{2x}\right)^N
  \left[ \left(
      C^{\psi}_{k,N}(a_s,\frac{Q^2}{\mu^2},\varepsilon) 
       Z^{\psi\psi}_N(a_s,\frac{1}{\varepsilon})
    + C^{G}_{k,N}(a_s,\frac{Q^2}{\mu^2},\varepsilon) 
       Z^{G \psi}_N(a_s,\frac{1}{\varepsilon})
  \right) \right. A^{\psi}_{{\rm quark},N}(a_s,\frac{p^2}{\mu^2},\varepsilon) 
\]
\begin{equation} \label{eqqaqa}
  \left. + \left(
      C^{\psi}_{k,N}(a_s,\frac{Q^2}{\mu^2},\varepsilon)
       Z^{\psi G}_N(a_s,\frac{1}{\varepsilon})
    + C^{G}_{k,N}(a_s,\frac{Q^2}{\mu^2},\varepsilon)
       Z^{G G}_N(a_s,\frac{1}{\varepsilon})
  \right)  A^{G}_{{\rm quark},N}(a_s,\frac{p^2}{\mu^2},\varepsilon) \right]
 + O(p^2)
\end{equation}
\[
 T^{ \rm g \gamma g \gamma}_{k}(p,q,a_s,\mu^2,\varepsilon)  = \hspace{13cm} \]
\[   \sum_{N=2}^{\infty} \left( \frac{1}{2x}\right)^N
  \left[ \left(
      C^{\psi}_{k,N}(a_s,\frac{Q^2}{\mu^2},\varepsilon)
       Z^{\psi G}_{N}(a_s,\frac{1}{\varepsilon})
    + C^{G}_{k,N}(a_s,\frac{Q^2}{\mu^2},\varepsilon)
       Z^{GG}_N(a_s,\frac{1}{\varepsilon})
    \right) A^{G}_{{\rm gluon},N}(a_s,\frac{p^2}{\mu^2},\varepsilon) \right.
\]
\begin{equation} \label{eqgaga}
  \left. + \left(
      C^{\psi}_{k,N}(a_s,\frac{Q^2}{\mu^2},\varepsilon)
       Z^{\psi\psi}_N(a_s,\frac{1}{\varepsilon})
    + C^{G}_{k,N}(a_s,\frac{Q^2}{\mu^2},\varepsilon)
       Z^{G \psi}_N(a_s,\frac{1}{\varepsilon})
  \right) A^{\psi}_{{\rm gluon},N}(a_s,\frac{p^2}{\mu^2},\varepsilon) \right]
 + O(p^2) \end{equation}
where  $k$ =2,L, $a_s \equiv a_s(\mu^2/\Lambda^2)$ 
 and it is understood that the l.h.s. is renormalized by 
substituting the bare coupling constant in terms of the renormalized one,
\begin{equation}
       a_b = a_s - \frac{\beta_0}{\varepsilon} a_s^2
          + (\frac{\beta_0^2}{\varepsilon^2}
         - \frac{\beta_1}{2 \varepsilon}) a_s^3 + O(a_s^4) 
\end{equation}
The terms $O(p^2)$ in the r.h.s. of Eqs. (\ref{eqqaqa}) and (\ref{eqgaga})
  indicate higher twist contributions.
The renomalization factors for the external quark and gluon lines 
are overall factors on both sides of the equations and are omited. 
The coefficient functions on the r.h.s are renormalized quantities.
The matrix elements $A^{i}_N$ are the  matrix elements of {\em bare} operators 
and are defined as in Eq. (\ref{mel}) with the nucleon states
replaced by the appropriate quark or gluon states.
\[ \vspace{10.5 cm} \]
\[ \parbox{14cm}{ {\bf Figure 1.}
A graphical representation of  Eqs. (\ref{eqqaqa}) and (\ref{eqgaga}).
The symbol $\otimes$ indicates an appropriate quark or gluon operator 
(defined in Eqs. (\ref{defoperatorns}) ,(\ref{defoperatorquark}), 
(\ref{defoperatorgluon})). On the l.h.s. of the equations also 
the crossed diagrams contribute but they are not explicitly shown. 
} \] \vspace{3mm} 

It is known that the gauge invariant operators $O^\psi$ and $O^G$ mix under
renormalization with unphysical operators (that are BRST variations
 of some operators or that vanish by the equations of motion)
\cite{collins,hn,scalise}.
But physical matrix elements (i.e. on-shell matrix elements with physical
polarizations) of such unphysical operators vanish.
Since the method  that is described below deals with
physical matrix elements we omited the unphysical operators in
Eqs. (\ref{eqqaqa}) and (\ref{eqgaga}).

Starting from  Eqs. (\ref{eqqaqa}) and (\ref{eqgaga}),
the anomalous dimensions and the coefficient 
functions are calculated using the method of projections of Ref. \cite{glt}.
It reduces the calculation of (moments of) coefficient functions and anomalous
dimensions to the calculation of diagrams of the propagator type 
instead of the 4-point diagrams that contribute to $T_{\mu\nu}$.  
This method relies heavily on the use of dimensional regularization and
the minimal subtraction scheme and implicitly involves a considerable 
rearrangement of infrared and ultraviolet divergences. 

The method consists of applying the following projection operator 
to both sides of Eqs. (\ref{eqqaqa}) and (\ref{eqgaga}).
\begin{equation} \label{projectionoperator}
 {\cal P}_N \equiv
  \left. \Biggl[ \frac{q^{ \{\mu_1}\cdots q^{\mu_N \}}}{N !}
  \frac{\partial ^N}{\partial
p^{\mu_1} \cdots  \partial p^{\mu_N}} \Biggr] \right|_{p=0} 
\end{equation}
Here $q^{ \{\mu_1}\cdots q^{\mu_N \}}$ is the harmonic (i.e.
symmetrical and traceless)
 part of the tensor $q^{\mu_1}\cdots q^{\mu_N }$ (see next section).
 The operator ${\cal P}_N$ is applied
to the integrands of all Feynman diagrams (nullifying $p$ before
taking the limit $\varepsilon \rightarrow 0$, to dimensionally regularize 
the infrared divergences
as $p\rightarrow 0$ for individual diagrams).
It is important to realize that this operation does not act on the 
renormalization constants $Z_N^{ij}$ and the 
coefficient functions on the r.h.s. 
of Eqs. (\ref{eqqaqa}), (\ref{eqgaga}). It does
however act on the matrix elements $A^{i}_N$. The nullification of
$p$ has the effect that of all the diagrams that contribute to the perturbative 
expansion of $A^{i}_N$ only the tree level terms (i.e. with no loops) 
survive since all diagrams containing loops become massless tadpole 
diagrams. Massless tadpole diagrams are put to 
zero in dimensional regularization.
Furthermore, the $N^{\rm th}$ order differentiation in the operator 
$ {\cal P}_N$ has the effect that ${\cal P}_N$ projects out only the 
$N^{th}$ moment since of all the factors $1/(2x)^{N'}$ only
 $1/(2x)^{N}$ gives a non zero contribution after nullifying $p$. 
On the left hand side the effect of ${\cal P}_N$ is
    to effectively reduce the 4-point diagrams that contribute
    to $T_{\mu\nu}$ to  2-point diagrams
    (this follows from the nullification of the momentum $p$),
    which drastically simplifies the calculation. 
We apply the operator ${\cal P}_N$ after the tensor structures $2,L$ have been 
projected out because the operator ${\cal P}_N$ would mutilate the tensor 
structure of $T_{\mu\nu}$. In the projector ${\cal P}_N$ we use the harmonic
tensor $q^{ \{\mu_1}\cdots q^{\mu_N \}}$ to remove higher twist contributions
(the $O(p^2)$ terms in Eqs. (\ref{eqqaqa}) and (\ref{eqgaga})) that after
differentiation with respect to  $p^{\mu}$ survive as terms proportional
to the metric tensor.

Summarizing, we have after application of the projection operator
$ {\cal P}_N$  to Eqs. (\ref{eqqaqa}) and (\ref{eqgaga})
\begin{equation} \label{qaqamom}
 T^{\rm q\gamma q\gamma}_{k,N}(\frac{Q^2}{\mu^2},a_s,\varepsilon)
     = \left(  C^{\psi}_{k,N}(a_s,\frac{Q^2}{\mu^2},\varepsilon) 
          Z_{N}^{\psi\psi}(a_s,\frac{1}{\varepsilon})
     + C^{G}_{k,N}(a_s,\frac{Q^2}{\mu^2},\varepsilon) 
  Z_{N}^{G \psi}(a_s,\frac{1}{\varepsilon})
             \right)     A_{{\rm quark},N}^{\psi,{\rm tree}}(\varepsilon)
\end{equation}
\begin{equation} \label{gagamom}
 T^{\rm g\gamma g\gamma}_{k,N}(\frac{Q^2}{\mu^2},a_s,\varepsilon) 
     = \left( C^{\psi}_{k,N}(a_s,\frac{Q^2}{\mu^2},\varepsilon) 
             Z_{N}^{\psi G } (a_s,\frac{1}{\varepsilon})
     + C^{G}_{k,N}(a_s,\frac{Q^2}{\mu^2},
       \varepsilon) Z_{N}^{GG} (a_s,\frac{1}{\varepsilon})
     \right)   A_{{\rm gluon},N}^{G,{\rm tree}}(\varepsilon) 
\end{equation}
where $k = 2,L$ and we defined
\begin{equation}
 T_{k,N}(\frac{Q^2}{\mu^2},a_s,\varepsilon) \equiv \left.
  \frac{ q^{ \{\mu_1}\cdots q^{\mu_N \}}}{ N !} \frac{\partial ^N}{\partial
 p^{\mu_1}\cdots  \partial p^{\mu_N}}T_k(p,q,a_s,\mu^2,\varepsilon)
 \right| _{p=0}. 
\end{equation}
It should be understood that (\ref{qaqamom}) and (\ref{gagamom})
    represent a large coupled system of equations
    when both sides are expanded in powers of $a_s$ and $\varepsilon$
    (i.e. $C$ is expanded in positive powers of $\varepsilon$ and Z is expanded
    in negative powers of $\varepsilon$).

After the calculation of $T^{\rm q\gamma q\gamma}$ 
and $T^{\rm g\gamma g\gamma}$ 
in the order $a_s^3$ and the determination of the tree level 
matrix elements 
$A_{N}^{j,{\rm tree}}$ one can solve Eqs. (\ref{qaqamom}) and (\ref{gagamom})
 simultaneously to obtain 
$C^{\psi}_{k}$, $C^G_k$, $Z^{\psi\psi}$ and $Z^{\psi G}$ in order $a_s^3$
but, unfortunately, $Z^{G \psi} $ and $Z^{GG}$ only in the order $a_s^2$.
This limitation follows directly from the fact that $C^{\psi}_{k}$ starts 
from the order $a_s^0$ but $C^G_{k} $ starts from order $a_s$ 
since the photon couples {\em directly} only to quarks. 
In solving the equations it is essential that all poles of the $Z$ factors are 
fully expressed in
terms of the anomalous dimensions as was discussed in the beginning of
this section. Coefficient functions and operator matrix elements are finite
as $\varepsilon \rightarrow 0$ but one must make sure that
sufficiently high powers in $\varepsilon$ are taken into account.
For example, one should consider order $\varepsilon^2$ contributions 
for $C^{\psi}_{k}$ at order $a_s$. We stress that by calculating only 
propagator type diagrams in the l.h.s. of Eqs. (\ref{qaqamom})
 and (\ref{gagamom})
we can get both renormalization constants of operators and coefficient
functions.

\[ \vspace{2.3cm}  \]
\[ \parbox{15cm}{ {\bf Figure 2.}
Examples for diagrams contributing to the Green functions
$T^{\rm g\phi g\phi}$ (a) and  $T^{\rm q\phi q\phi }$ (b).
} \] \\

  To obtain $Z^{G\psi}$ and $Z^{GG}$ in order $a_s^3$ we calculated
    two more unphysical Green functions $T^{\rm q\phi q\phi }$ and
    $T^{\rm g\phi g\phi}$ (see fig 2),
    in which the photon is replaced by an external scalar particle $\phi$
    that couples {\em directly} only to gluons.
    The vertices that describe the coupling between the external scalar field
    $\phi$ and the gluons follow from adding the simplest gauge invariant
    interaction term $\phi G_{\mu\nu}^{a}G^{\mu\nu}_a$
    (where $G_{\mu\nu}^{a}$ is the QCD field strength tensor)
    to the QCD Lagrangian.
    For the Green functions $T^{\rm q\phi q \phi}$ and $T^{\rm g\phi g \phi}$
    an OPE similar to (\ref{OPE}) exists with the same
    operators but with different coefficient functions $C^{G}_{\phi}$ and
    $C^{\psi}_{\phi}$,
    where $C^{\psi}_{\phi}$ starts
    from the order $a_s$ and $C^{G}_{\phi}$ starts from the order $a_s^0$.

Repeating the steps that led to Eqs. (\ref{qaqamom}) and (\ref{gagamom})
 one finds for these Green functions the following equations
\begin{equation} \label{gjgjmom}
 (Z_{G^2})^2 T^{\rm g \phi g \phi}_{N}(\frac{Q^2}{\mu^2},a_s,\varepsilon)
     = \left(
   C^{G}_{\phi,N}(a_s,\frac{Q^2}{\mu^2},\varepsilon)
  Z_N^{GG}(a_s,\frac{1}{\varepsilon})
     + C^{\psi}_{\phi,N}(a_s,\frac{Q^2}{\mu^2},\varepsilon) 
                       Z_N^{\psi G}(a_s,\frac{1}{\varepsilon})
            \right) A_{{\rm qluon},N}^{G,{\rm tree}}(\varepsilon)
\end{equation}
\begin{equation} \label{qjqjmom}
 (Z_{G^2})^2 T^{\rm q \phi q \phi}_{N}(\frac{Q^2}{\mu^2},a_s,\varepsilon)
     = \left(  C^{G}_{\phi,N}(a,\frac{Q^2}{\mu^2},\varepsilon) 
                       Z_N^{G \psi} (a_s,\frac{1}{\varepsilon})
     + C^{\psi}_{\phi,N} (a_s,\frac{Q^2}{\mu^2},\varepsilon) 
                            Z_N^{\psi\psi} (a_s,\frac{1}{\varepsilon})
                  \right)  A_{{\rm quark},N}^{\psi,{\rm tree}}(\varepsilon) 
\end{equation}
As is indicated in the l.h.s., the external operators 
$G_{\mu\nu}^{a}G^{\mu\nu}_a$  have to be renormalized \cite{kluberg}
\[ \left(G_{\mu\nu}^{a}G^{\mu\nu}_a\right)_{R} = Z_{G^2} 
\left(G_{\mu\nu}^{a}G^{\mu\nu}_a\right)_{B}+\cdots, \hspace{2cm} 
 Z_{G^2} = \frac{1}{1-\beta (a_s)/( a_s \varepsilon)} \]
where the dots indicate (unphysical) operators that are omited since
they do not contribute to the on-shell matrix elements with physical
spin projections that we consider. (The only physical operator that
mixes under renormalization with $G_{\mu\nu}^{a}G^{\mu\nu}_a$ is the
quark operator $m_q \overline{\psi} \psi$ that vanishes in our limit
of massless quarks.)
We emphasize that the 3-loop $\beta$-function is required
for the present 3-loop calculation.

After the calculation of $T^{\rm q\phi q\phi}$  and
$T^{\rm g\phi g\phi}$ in the order $a_s^3$ 
one can solve Eqs. (\ref{gjgjmom}) and (\ref{qjqjmom})  to obtain
$C^{G}_{\phi}$, $C^{\psi}_{\phi}$, $Z^{G\psi}$ and $Z^{GG}$ in the order $a_s^3$
(from these equations $Z^{\psi G}$ and $Z^{\psi\psi}$ can be obtained
in the order $a_s^2$ only). 
Please note that the coefficients $C^{G}_{\phi}$ and
 $C^{\psi}_{\phi}$ are obtained 
as a byproduct and are not important for the physical process 
under consideration. 
Furthermore, the two sets of equations (\ref{qaqamom}), (\ref{gagamom}) and
  (\ref{gjgjmom}), (\ref{qjqjmom}) determine all the anomalous
dimensions of the order $a_s^2$ in two independent ways which provides 
a consistency check on the results.

Solving the system of equations for $F_L$ is slightly more involved
than for $F_2$ since the structure function $F_L$ contains no tree level 
(i.e. order $a_s^0$) contribution.
To solve the sets of equations for  $F_L$ one should add extra information,
for example the ${\cal O}(a_s^2)$ contributions to $Z^{G\psi}$ and $Z^{GG}$
as determined from the equations for $F_2$.

\section{The calculation}

As was discussed in the previous section, we will apply the
operator ${\cal P}_N$ for $N$=2,4,6,8 to 4 different Green functions,
$T_{k}^{\rm q\gamma q\gamma}$,  $T_{k}^{\rm g\gamma g\gamma}$, 
$T^{\rm q\phi q\phi}$ and $T^{\rm g\phi g\phi}$ 
and we sum over the physical spin polarizations of the external 
quarks and gluons.
For the external quarks (in $T^{\rm q\gamma q\gamma}$ 
and $T^{\rm q\phi q\phi}$) the sum over the polarizations is performed 
by inserting the projection operator $\not\hspace{-1mm} p$ 
between the external quark legs
and taking the trace over the strings of gamma matrices.
For the external gluons (in $T_{k}^{\rm g\gamma g\gamma}$ 
 and $T^{\rm g\phi g\phi}$) 
the sum over physical spins can be done by contracting the external 
gluon lines with $-g^{\alpha\beta}+(p^\alpha q^\beta +p^\beta q^\alpha)/
p\cdot q - p^\alpha p^\beta q^2/(p\cdot q)^2$ in which the (on-shell) gluon
has momentum $p$ (with $p^2 = 0$). The presence of the extra powers of $p$
poses considerable efficiency problems (the operator ${\cal P}_N$ will
generate more than 3 times larger intermediate expressions as compared 
to the case of a simpler $g^{\alpha\beta}$ projection).  
Alternatively one may take the sum over physical gluon spins by
contracting the external gluon lines with only $-g^{\alpha\beta}$ and adding
external ghost contributions to the Green functions,
$T_{k}^{\rm h\gamma h\gamma}$ to $T_{k}^{\rm g\gamma g\gamma}$ and 
$T^{\rm h\phi h\phi}$ to $T^{\rm g\phi g\phi}$, 
where the label h indicates an external ghost line.
This procedure is identical to the standard use of ghost diagrams to remove  
unphysical polarizations of gluon propagators in the covariant gauge, and the
ghost particle h is the same ghost that we use in closed loops.
Although we now have to consider all diagrams that contribute to
$T_{k}^{h\gamma h\gamma}$ 
and $T^{\rm h\phi h\phi}$, and increase the total number of
diagrams that we have to calculate, it still makes the computations
more than a factor of 3 faster (since ghost diagrams are of a far simpler 
nature than gluon diagrams). We checked for the lowest moments
that the two methods for taking the sum over the gluon spin polarizations
gave the same results, but for the higher moments we only applied the ghost
method. 
\[ \vspace{2.3cm} \]
\[ \parbox{15cm}{ {\bf Figure 3.}
Examples for ghost diagrams contributing to the Green functions
$T^{\rm h\gamma h\gamma}$ (a) and  $T^{\rm h\phi h\phi }$ (b)
} \] \\

The explicit generation of Feynman diagrams with the corresponding
symmetry factors has been done automatically by the use of the program
QGRAF\cite{qgraph}.
 Statistics on the number of diagrams in the different classes
q$\gamma$q$\gamma$, q$\phi$q$\phi$ etc. is presented in table 1. 
The generation (and the counting) of the diagrams
    is non-standard, since a number of tricks have been used
    (for example, crossed diagrams are not generated explicitly,
     and the diagrams with 4-gluon vertices are split into
     a number of parts for which the colour factor necessarily
     factorizes).

\[ \begin{tabular}{l c c c c c}\\
\hline
{} &{Tree} &1-loop &2-loops &3-loops &Lorentz projections\\
\hline
q$\gamma$q$\gamma$
   &1  & \phantom03  & \phantom027           & \phantom0\phantom0413 &2\\
q$\phi$q$\phi$ 
     &{} & \phantom01  & \phantom024           & \phantom0\phantom0697 &1\\
g$\gamma$g$\gamma$
     &{} & \phantom02  & \phantom020           & \phantom0\phantom0366 &2\\
h$\gamma$h$\gamma$
     &{} & {}          & \phantom0\phantom02   & \phantom0\phantom0\phantom053 &
2\\
g$\phi$g$\phi$ &1  & 11          & 241                   &  \phantom07219 &1\\
h$\phi$h$\phi$ &{} & \phantom01  & \phantom036           &  \phantom01266 &1\\
\hline
TOTAL &3 & 23          & 399                   & 10846 &{}\\
\hline\\
\end{tabular} \]
\[ \parbox{14cm}{ {\bf Table 1.}
Number of diagrams and Lorentz tensor structures
  in the classes q$\gamma$q$\gamma$, q$\phi$q$\phi$, g$\gamma$g$\gamma$,
   h$\gamma$h$\gamma$, g$\phi$g$\phi$ and h$\phi$h$\phi$.
Notation: q = quark, g = gluon, h = ghost, $\gamma$ = photon, $\phi$ =
scalar particle that couples only to gluons.
} \]
It is clear from these statistics that the calculation of the diagrams
necessarily has to be automated to a large extent. The calculation is
therefore organized as follows:
\begin{enumerate}
\item The diagrams are generated automatically with a special version of
the diagram generator QGRAF\cite{qgraph}.
For every class q$\gamma$q$\gamma$, q$\phi$q$\phi$ etc.
the full set of diagrams is put into a single file
 using a dedicated database program MINOS that manages information
about thousands of diagrams and can be instructed to call other programs,
giving them the proper information from a database file.
\item
The representation for a diagram at this point is still a very compact one,
and this is explored as follows.
We use programs written for the symbolic manipulation
program FORM\cite{form} to calculate colour factors for each diagram
and bring the diagrams into a representation that explicitly contains
information required at later steps in the calculation.
For instance, this involves choosing automatically an optimal path 
(in most cases the shortest path) for the external
momentum $p$ to flow through each diagram (we are going to expand in $p$
when the operator ${\cal P}_N$ is applied) and determining automatically the
diagram's topology when $p$ is nullified.
This information, for all diagrams together, is kept in a single file 
and is accessible using MINOS.
\item  We instruct MINOS to run sequentially, one diagram at a time,
  a highly optimized FORM program that performs the explicit calculation
  i.e. it substitutes all the Feynman rules,
it performs projections on the Lorentz structures of the Green functions,
  it Taylor-expands the diagram in the external momentum $p$ 
(the depth of the expansion increases with the moment index N),
it takes all the Dirac traces, contracts with the tensors
 $q^{ \{\mu_1}\cdots q^{\mu_N \}}$
 and finally calls the MINCER \cite{mincer} 
integration package to perform the 3-loop scalar
integrals of the massless propagator type (using the integration by parts
algorithms published in Ref. \cite{ibp}). 
The results together with some useful technical information 
about the calculation (such as the resources used) are again stored 
into a single file. MINOS will initiate the calculation of a next diagram
as soon as the calculation of a previous diagram is completed
without any need for human interference.

\end{enumerate}

An important aspect of the calculation is the use of 
the symmetrical and traceless tensors, $q^{ \{\mu_1}\cdots q^{\mu_N \}}$
that are used to extract the leading twist contributions.
These tensors are (with the proper normalization)
known as `harmonic tensors' $H_{n}$ satisfying 
\[
\begin{array}{llll}
 H_{n}^{\mu_1\cdots\mu_n}\delta^{\mu_1\mu_2}  &  = & 0 &
   {\rm (traceless)}  \\
 H_{n}^{\mu_1\cdots\mu_i\cdots\mu_j\cdots\mu_n} & = &
  H_{n}^{\mu_1\cdots\mu_j\cdots\mu_i\cdots\mu_n} &
   {\rm (symmetrical)}  \\
 H_{n}^{\mu_1\cdots\mu_n}Q^{\mu_n} & = & H_{n-1}^{\mu_1\cdots\mu_{n-1}} Q^2
  & {\rm (normalization)}
\end{array}
\]
and an explicit construction (in Euclidean space-time)
in terms of the Kronecker delta symbols
$\delta^{\mu\nu}$ and the momenta $Q^{\mu}$ reads
\[ H_{n}^{\mu_1\cdots\mu_n} =
    \sum_{j=0,2,\cdots,n} h_j^n \left(Q^2\right)^{j/2} \sum_{\rm index 
\hspace{1mm} perm.}
   \delta(\mu_1,\cdots,\mu_{j}) Q^{\mu_{j+1}}\cdots Q^{\mu_n} ,\]
\begin{equation} \label{harmo}
 h_j^n = (-1)^{j/2} 2^{n-j/2}
  \frac{ \Gamma(2-2\varepsilon) 
         \Gamma(1-\varepsilon +n-j/2)}{
        \Gamma(2-2\varepsilon+n) \Gamma(1-\varepsilon) } 
\end{equation}
where the second summation is over all the ways to partition  
the indices into one set containing $j$ indices (put into $\delta$ )
 and a second set 
containing $(n-j)$ indices (put on momenta $Q\cdots Q$).
$\Gamma$ is the Euler gamma function (factorial function).
 The tensor $\delta$ is a completely 
symmetrical tensor constructed from Kronecker delta symbols only,  
\[
\begin{array}{lll}
\delta(\mu_1,\mu_2) & = &\delta^{\mu_1\mu_2},\\
 \delta(\mu_1,\mu_2,\mu_3,\mu_4) & = &\delta^{\mu_1\mu_2} \delta^{\mu_3\mu_4}
 + \delta^{\mu_1\mu_3} \delta^{\mu_2\mu_4} +
\delta^{\mu_1\mu_4} \delta^{\mu_2\mu_3},\\ {\rm  etc.} & & 
\end{array} \]
such that the normalization is
\[ \delta(\mu_1,\mu_1,\mu_2,\mu_2,\cdots,\mu_n,\mu_n) = 2^n \frac{
 \Gamma (2-\varepsilon +n) }{\Gamma (2-\varepsilon ) } \]
For example the tensor with four indices is 
\[ \begin{array}{lll}
 H_{4}^{\mu_1,\mu_2,\mu_3,\mu_4} & =  &
 h_4^4 Q^4 \biggl[ \delta^{\mu_1\mu_2} \delta^{\mu_3\mu_4}
 + \delta^{\mu_1\mu_3} \delta^{\mu_2\mu_4}
 + \delta^{\mu_1\mu_4} \delta^{\mu_2\mu_3} \biggr] 
 + h_0^4 Q^{\mu_1} Q^{\mu_2} Q^{\mu_3} Q^{\mu_4} \\
 & & + h_2^4 Q^2 \biggl[  \delta^{\mu_3\mu_4} Q^{\mu_1} Q^{\mu_2} 
                    + \delta^{\mu_2\mu_4} Q^{\mu_1} Q^{\mu_3}
                    + \delta^{\mu_2\mu_3} Q^{\mu_1} Q^{\mu_4} \\
 & &                + \delta^{\mu_1\mu_4} Q^{\mu_2} Q^{\mu_3}
                    + \delta^{\mu_1\mu_3} Q^{\mu_2} Q^{\mu_4}
                    + \delta^{\mu_1\mu_2} Q^{\mu_3} Q^{\mu_4} \biggr] .
\end{array} 
\]
The number of terms in a harmonic tensor increases rapidly with the
rank of the tensor and in the present calculation these harmonic tensors 
are contracted with many other
tensors constructed from only a few different (integration) momenta.
An efficient implementation of the harmonic tensors that directly takes 
into account the symmetries of the other tensors 
can greatly limit the number of terms that are produced in the calculation
and is vital for its 
feasibility\footnote{it is therefore interesting to mention that an efficient
 implementation of Eq. (\ref{harmo}) exists in FORM-2 and requires only 
 3 lines of code using the built in combinatorical functions.}.

The results of our calculations are presented in the last section of
this article.
The results for the second moment N=2, that are presented in 
this paper, were obtained in an arbitrary
covariant gauge for the gluon fields. This means that we keep 
the gauge parameter $\xi$ that appears in the gluon propagator 
$ i$ $[-g^{\mu\nu}+(1-\xi)
 q^{\mu}q^{\nu}/(q^2+i\epsilon)]/(q^2+i\epsilon)$ as a
free parameter in the calculations. The explicit cancellation of the gauge 
dependence in the coefficient functions and anomalous dimensions
gives an important check of the results. 

Another non-trivial check of the calculation is the validity of the 
singlet relations $ \gamma_2^{\psi \psi} + \gamma_2^{G \psi} = 0 $,
$ \gamma_2^{\psi G} + \gamma_2^{G G} = 0 $ that follow from the
fact that the QCD energy momentum tensor (which contains both the spin-2 
quark and gluon operators) is conserved and has therefore 
zero anomalous dimension. 

Please note that at present the only independent check of our
3-loop results  is provided by the calculation
in Ref.\cite{gracey} where the leading $n_f$ terms were calculated
for the non-singlet anomalous dimensions in all orders in $a_s$ using
a large $n_f$ expansion. Our $a_s^3 n_f^2$ terms agree with Ref.\cite{gracey}.

The reader may notice the appearance of the constant $\zeta_4$ in the final 
3-loop results for the coefficient functions 
$C^{\psi}_{2,N}$ and $C^{G}_{2,N}$ which seems to be in 
conflict with the empirical law that in the results for inclusive
physical quantities $\zeta_4$ does not appear. However
these 3-loop coefficient functions contribute to the next-to-NNL (NNNL) order
for $F_2$ and require 4-loop anomalous dimensions to get a complete
physical NNNL approximation. Assuming the cancellation of $\zeta_4$
in the complete NNNL approximation, one can derive the coefficients
of $\zeta_4$ in the 4-loop anomalous dimensions.

In spite of all optimizations of the integration program MINCER and of many
other efficiency-crucial parts of the calculation, 
the calculation of the singlet moments, $N$=2,4,6,8 as published in this
paper, required more than the equivalent of 7500 hours on a 150 Mhz SGI
Challenge workstation and required at some instances 2 Gbyte
of storage place for the intermediate stages in the calculation of one of
   the diagrams\footnote{This is called `intermediate expression swell'. It is
   a well known phenomenon in Computer Algebra.}.
We noticed that both the required disk storage space and computation time
increased with almost a factor 5 when we compared the calculation of each 
(N+2)th moment to the calculation of the corresponding Nth moment.
The calculation of higher moments using the same methods is therefore
not feasible at present.

\section{Analysis}
In this section we will investigate the effects due to the 
calculated 3-loop coefficient 
functions and anomalous dimensions (the list of the
analytical 3-loop results is given in the next section).
We will reconstruct the
coefficient functions and splitting functions as distributions in
$x$-space using the calculated moments and incorporating as much as possible 
the known information
about the leading (and in some cases the next-to-leading) singularities
in the limits $x\rightarrow 1$ and $x\rightarrow 0$. Explicit
comparison with the known 2-loop coefficient
functions and anomalous dimensions
 shows that our approach to the reconstruction gives very good 
effective fits of the $x$-space distributions when only few moments 
are used and when sufficient information about the two endpoints $x=0$,
$x=1$ is known. 
This procedure also allows us to estimate 
the error of the reconstruction.

In order to relate the calculated results for the
anomalous dimensions $\gamma_{N}$  and the coefficient functions $C_N$ 
to experiment one must be able to obtain
the experimentally measurable structure functions $F_k(x,Q^2)$
from the Mellin moments $M_{k,N}$ of Eq. (\ref{Mkn}).
The rigorous procedure to obtain $F_k(x,Q^2)$
from the moments is to apply the inverse Mellin transform 
which, however, requires the exact knowledge of $M_{k,N}$ in the
complex $N$-plane (or equivalently the analytic continuation to complex
$N$ from all even or odd moments). 
Because at the 3-loop order we have calculated only a limited number 
of moments  $\gamma_{N}$ and $C_N$, we can obtain only 
approximate results in $x$-space. An example of a NNL order analysis 
in $x$-space based on a limited number of non-singlet moments
can be found in Ref. \cite{anakotikov}.

As an alternative to studying the $Q^2$-dependence of
$F_k(x,Q^2)$ via the $Q^2$-dependence of moments $M_{k,N}$ 
(see Eq. (\ref{rg})), one may start from the $x$-space distributions   
$\gamma (x) $ and $C (x)$ that are related to $ \gamma_{N}$
and $ C_N$ via the Mellin transform
\begin{equation}
 \gamma_{N}  = \int_0^1 dx\hspace{1mm} x^{N-1} \gamma (x) ,\hspace{.5cm}
 C_N  =  \int_0^1 dx\hspace{1mm} x^{N-1} C (x)
\end{equation}
and do the $Q^2$-evolution in $x$-space via the DGLAP equations \cite{glap}

\begin{equation} \label{ap1}
Q^{2}\frac{d}{dQ^{2}}\left(
\begin{array}{l} q^{+}(x,Q^{2}) \\ G(x,Q^{2})
\end{array} \right) = - \left( \begin{array}{ll}
\gamma^{\psi\psi}(x,a_s) & \gamma^{\psi G}(x,a_s) 
\\ \gamma^{G\psi}(x,a_s) & \gamma^{GG}(x,a_s)
\end{array} \right) \otimes \left( \begin{array}{l} q^{+}(x,Q^{2}) \\
G(x,Q^{2}) \end{array} \right) \end{equation}

\begin{equation} \label{ap2}
 Q^{2}\frac{d}{dQ^{2}}\left(q^{+}_{f}(x,Q^{2})\!-\!\frac{q^{+}(x,Q^{2})}{n_f}
\right)\!=\!- \gamma^{ns,+}(x,a_s)
\otimes\! \left(q^{+}_{f}(x,Q^{2})\!-\!\frac{q^{+}(x,Q^{2})}{n_f}
\right) \end{equation}

\begin{equation} \label{ap3} Q^{2}\frac{d}{dQ^{2}}
 \left( q^{-}_{f}(x,Q^{2})\right) = - \gamma^{ns,-}(x,a_s)
\otimes q^{-}_{f}(x,Q^{2}) \end{equation}
where
$q_f^{\pm} = q_f\pm \overline{q_f}$, $q^{+} = \sum_{f=1}^{n_f}q_f^{+}$,
$q_f \in \{ u, d, s, c, t, b\}$ are the quark distributions and $G$ is the
gluon distribution. It is understood that the inverse Mellin transforms
of the anomalous dimensions are related to the standard 
Altarelli-Parisi
splitting functions $P^{ij}(x)$ as $\gamma^{ij} (x) = - P^{ij}(x)$.
 The symbol $\otimes$ indicates the convolution integral
\begin{equation}
A(x)\otimes B(x) \equiv \int_{0}^{1}dx_{1} \int_{0}^{1} dx_{2}A(x_{1})
B(x_{2}) \delta(x-x_{1}x_{2}) = \int_{x}^{1}\frac{dx_{1}}{x_{1}}
A(\frac{x}{x_{1}}) B(x_{1}). \end{equation}
Since we calculated the moments only for electromagnetic scattering
we are restricted to the ``+'' type anomalous dimensions 
and coefficient functions for the
non-singlet sector, i.e. to the distributions that correspond to
analytic continuation from the even moments
(i.e. we do not consider the ``-'' type distributions which
correspond to analytic continuation from the odd moments and
which are relevant for processes that are not symmetric with respect to
crossing symmetry such as neutrino scattering via $W$-boson exchange).
The proper relation between the structure functions and 
parton distributions for electromagnetic scattering is
\begin{eqnarray}
 F_k(x,Q^2)& = & x \left\{ \sum_{f=1}^{n_f}  e_f^2
 \Biggl[  C_k^{ns,+}(x,a_s)\otimes  
  \left(q^{+}_{f}(x,Q^{2})\!- \frac{1}{n_f}
  q^{+}(x,Q^{2}) \right) \Biggr] \right. 
 \nonumber \\
& & \left. + \frac{1}{n_f} \sum_{f=1}^{n_f} e_f^2 
   \Biggl[     C_k^{\psi}(x,a_s)\otimes q^{+}(x,Q^{2})
           + C_k^{G}(x,a_s)\otimes G(x,Q^{2}) \Biggr] \right\}, 
\end{eqnarray}
where $e_f$ is the electromagnetic charge of a quark of the flavour $f$.

A special feature of the $Q^2$-evolution in $x$-space (that is a direct
consequence of the convolution integral) is that the
structure functions $F_k(x,Q^2)$ at some value $x=x_0$ only depend 
on the parton distributions, splitting functions
and coefficient functions in the region $ x_0 \leq x \leq 1$. 
This means that one does not need to know the anomalous dimensions 
and coefficient functions below the experimentally accessible $x$-region
(but one always needs them for $x \approx 1$).

One can see that, away from the singular limits
$x\rightarrow 0$ and $x\rightarrow 1$
that received special attention in the literature,
the 1-loop and 2-loop $\gamma (x)$ and $C (x)$  
\cite{grwilcz}-\cite{lvm} behave smoothly 
(as a smooth interpolation between the small-$x$ and large-$x$ regions).
As a consequence the distributions  $\gamma (x)$ and $C (x)$ can be
approximated to a high precision by linear combinations 
of simple distributions
that contain the singular\footnote{
$[f(x)]_+$ type distributions appear below, where $f(x)$ is non-integrable at 
$x = 1$.
The standard definition of integrals over these distributions 
is: $\int_a^1 g(x) [f(x)]_+ dx 
 = \int_0^1 \left( g(x)-g(1)\right) f(x) dx - \int_0^a g(x)f(x)dx $.
Or in a more symbolic form $[f(x)]_+ = f(x) - \delta(1-x) \int_0^1 f(z)dz$. }
 terms of the expansions of the exact $\gamma (x)$ and $C (x)$
in the limits $x\rightarrow 0$ and $x \rightarrow 1$,
i.e. $\log^i(x)/x$ and $\log^i(x)$ for small $x$ and  
$ [ \log^i(1-x)  /(1-x)]_{+} $, $\log^i(1-x)$ and $\delta(1-x)$
for large $x$ 
plus finite order polynomials of $x$ 
(the order determines the accuracy of the approximation).
(The exact results for $\gamma (x)$ and $C (x)$
contain in general complicated polylogarithmic functions but after
their expansions in the small-$x$ and large-$x$ regions one finds
the simple structure of the singularities mentioned above.)

The most singular terms of $\gamma (x)$ and $C (x)$ in the limit 
$x\rightarrow 0$ 
(i.e. $\log^i(x)/x$ ) show up as singularities in moment space 
for $N = 1$. By increasing the moment index $N$, 
one increasingly probes the region $x \approx 1 $ (since the factor $x^{N-1}$
suppresses the contribution from the small-$x$ region).
 In particular, when the moments
grow for large N this is due to terms of the type  
 $ [ \log^i(1-x)  /(1-x)]_{+} $, $i \geq 0$
(of which the moments increase as $ \log^{i+1}(N) $ for large N).

The approximations of $\gamma (x) $ and $C (x)$ in terms of 
the linear combinations of simple distributions (i.e. $\log^i(x)/x$,
$\log^i(x)$, $[ \log^i(1-x) /(1-x) ]_{+}$, 
 $\log^i(1-x)$, $\delta(1-x)$  and low powers of $x$ )
provides a natural scheme to study the effects of the 3-loop
moments N=2,4,6,8,(10)
on the $Q^2$-dependence of structure functions in $x$-space
and to quantify the uncertainty due to our lack of 
knowledge of the higher (i.e. $N > 10$) 3-loop moments.
Considering such approximate distributions for 
the a priori unknown $\gamma (x) $ and $C (x)$ at 3-loops
we may impose the correct asymptotic moment behavior 
for $N\rightarrow \infty$ and $N \rightarrow 1$ 
by including in the linear combinations of simple distributions
the correct leading (and subleading) 
terms in the large-$x$ and small-$x$ regions.
In some cases  information about 
these leading (and subleading) terms in the large
and small $x$ regions is available in the literature.

The linear combinations of distributions should also include various functions 
of a less singular type (compared to the leading singularities),
e.g. $\log^i(x)$, $\log^i(1-x)$ and low 
powers of $x$, which are relevant terms in the intermediate $x$ region.
We fit all unknown coefficients in a linear combination of simple
distributions such that the moments N=2,4,6,8,10 of the linear combinations
 give the calculated values. 
Since we know only 4 singlet (and 5 non-singlet) moments explicitly 
we can only allow
up to 4 (or 5) arbitrary coefficients (i.e. types of simple distributions) 
in the linear combinations. 
 Only in the case of the quark anomalous dimension we can
 allow more than 5 arbitrary coefficients
 by considering pure-singlet (ps) and non-singlet (ns) parts 
 separately (since information about the small-$x$ and large-$x$ behaviour
is known for these parts separately). 

Although at the 3-loop order there are only a finite number 
of allowed simple distributions of the types 
$\log^i(x)/x$, $[ \log^i(1-x) /(1-x) ]_{+}$, 
$\delta(1-x)$, $\log^i(x)$, $\log^i(1-x)$ 
we do not have enough moments to include all of them simultaneously in a fit
and we have to choose among the possible simple distributions.
Therefore, to obtain an estimate of the uncertainty of the reconstructions 
due to our lack of knowledge of the higher moments
we studied the stability of the $x$-space reconstructions  with respect 
to a change of the simple distributions included in the linear
combinations. If one does not find stable results under such a change
one should have included a larger number  
of simple distributions of the types mentioned above. 
However, as one can see below, in many cases  
we have sufficient information to obtain stable results.

For a correct $x$-space reconstruction it is crucial that one knows 
the moments exactly (or with a high precision) since the inversion of 
moments into $x$-space distributions from a limited set of moments 
 is numerically unstable,
i.e. a small change in one of the moments can lead to a large change 
in the reconstructed result. 
In relation to this, we note that also the dedicated
procedure for reconstructing $\gamma (x)$ and $C (x)$ that  
we consider here  
requires very accurate numerical values for the moments.
Therefore one can not rely on interpolation between known even moments 
to increase the number of available moments.  
\subsection{Anomalous Dimensions}

We will now consider the 3-loop singlet and non-singlet anomalous
dimensions. We will use the following notation 
for the inverse Mellin transformed anomalous dimensions
\begin{eqnarray} \gamma^{ns}(x,a_s)& = & a_s\gamma^{ns,(0)}(x)  +
     a_s^2 \gamma^{ns,(1)}(x)  + a_s^3 \gamma^{ns,(2)}(x) + O(a_s^4) \\
 \gamma^{ij}(x,a_s) & = & a_s \gamma^{ij,(0)}(x) +
                  a_s^2 \gamma^{ij,(1)}(x) + a_s^3 \gamma^{ij,(2)}(x) 
+ O(a_s^4), \hspace{1cm} i,j = \psi,G
 \end{eqnarray}
where $a_s = \alpha_s/(4\pi) $.
In Ref.\cite{cathautns} it was proven that non-singlet 
anomalous dimensions are finite 
as $N \rightarrow 1$ which means that in $x$-space the non-singlet
anomalous dimensions are less singular than $1/x$ as $x\rightarrow 0$.
The leading singularity has been  derived \cite{buemvogt} from 
 the leading order small-$x$ resummation of the non-singlet evolution
kernels \cite{kirslip} 
\begin{equation} \gamma^{ns,(2)}(x\rightarrow 0)  \label{leadns}
= -\frac{2}{3}C_F^3\log^4(x) + O(\log^3(x)). \end{equation}
The singlet anomalous dimensions are in general singular as
    $N\rightarrow 1$ ( ie $x\rightarrow 0$ ).
    The following results, which have been derived using
    small-$x$ resummations, may be found in the literature
    \cite{kurlipfad,basciamar,cathautns}.
\begin{eqnarray} \label{leads}
 \gamma^{\psi\psi,(2)}(x\rightarrow 0) & = & \frac{896}{27}n_f C_A C_F\log(x)/x 
     + O(1/x) \nonumber \\
 \gamma^{\psi G,(2)}(x\rightarrow 0) & = & \frac{896}{27} n_f C_A^2\log(x)/x 
     + O(1/x) \nonumber \\
 \gamma^{G \psi,(2)}(x\rightarrow 0) & = & O\left(\log(x)/x\right)
  \nonumber \\
 \gamma^{GG,(2)}(x\rightarrow 0) & = & O\left(\log(x)/x\right).  
\end{eqnarray}
The last two equations simply mean that there are no leading
singularities of the type
 $ \log^2 (x)/x $ (since the $a_s^3$ term in the BFKL anomalous dimension 
\cite{kurlipfad} vanishes).

For the limit $x\rightarrow 1$ of the anomalous dimensions not much is
known except for the conjecture \cite{gly} 
that the diagonal elements $\gamma^{\psi\psi}$ and
$\gamma^{GG}$ do not rise faster than $\log( N) $ as 
$ N\rightarrow \infty$, which means that one has in $x$-space  a leading term 
$ [ 1 /(1-x) ]_{+} $ (and not $[\log(1-x)/(1-x) ]_{+}$). 
This conjecture is based on
the explicitly known one and two loop results. But the rise 
of the 3-loop non-singlet moments observed for $N=2,4,6,8,10$ 
indicates that it also holds at the 3-loop level 
(putting both the terms $[1/(1-x)]_{+}$ and $[\log(1-x)/(1-x) ]_{+}$ 
in the linear combination of simple distributions gives a
small numerical coefficient for the second term compared to the first
one, although this is clearly not a proof).
Furthermore, the one and two loop coefficients of the terms 
$[ 1/(1-x) ]_{+}$ in $\gamma^{\psi\psi}$ and $\gamma^{GG}$ 
have a simple ratio $C_F/C_A$ and we presume that this ratio is the same 
at 3-loops.

The off-diagonal singlet anomalous dimensions $\gamma^{\psi G}$ and
$\gamma^{G\psi}$ go to zero as $N \rightarrow \infty$.
The best way to see that they do not contain terms
$ [ \log^i(1-x) /(1-x) ]_{+} $, $i\geq 0$,
is to consider the approach of Refs. \cite{grwilcz,frs,hn} 
to  calculate the anomalous dimensions, 
 i.e. to  study the renormalization of the singlet
twist-2 operators where one can see that the terms
$ [ \log^i(1-x)  /(1-x) ]_{+} $ originate from diagrams that 
only appear for diagonal anomalous dimensions (notice that this method
allows the direct calculation of these terms without performing the
complete calculation of the anomalous dimensions).
Therefore, the leading terms in $\gamma^{\psi G}$ and
$\gamma^{G\psi}$ as $x \rightarrow 1$ that one can expect from performing 
the necessary integrals at 3-loops are $\log^{4}(1-x)$. 

Finally we presume that the pure singlet (ps) combination 
$\gamma^{ps} = \gamma^{\psi\psi} - \gamma^{ns}$ 
rapidly vanishes as $N\rightarrow \infty$ (i.e. vanishes at least as quick
as $1/N$ for large $N$; at 2-loops it is known to vanish as $1/N^4$).  
We have observed this tendency already in the low-N moments
 also at 3-loops. 
This means that $\gamma^{\psi\psi}$ and $\gamma^{ns}$ contain the same 
terms that are important for $x\sim 1$ 
(since the transform from $x$- to N-space for large N
 gives $ [ \log^i(1-x)  /(1-x) ]_{+}\rightarrow \log^{i+1}(N) $, 
 $\delta(1-x)\rightarrow 1$ and $\log^i(1-x)\rightarrow \log^i(N)/N$ ) and
that they only differ in terms that are important for small-$x$.
In relation to this, recall that only $\gamma^{\psi\psi}$ contains a term 
$\log(x)/x$ and $\gamma^{ns}$ does not, see Eqs. (\ref{leadns},\ref{leads}). \\

\[ \vspace{15cm} \]
\[ \parbox{14cm}{{\bf Figure 4.}
The exact 2-loop anomalous dimensions and reconstructions
(approxs.) based on the lowest 5 moments N=2,4,6,8,10 ($n_f$ = 4).
The approximations are obtained by fitting the sets of 
distributions of Eq. (\ref{ano2loop}) to these 5 moments } \]

\newpage
At 2-loops the analysis based on only a few low-N anomalous dimensions
can be compared with the exact results as is shown in figure 4. 
The indicated approximations in figure 4 are based on matching 
the  linear combinations of simple distributions of Eq. (\ref{ano2loop})
to 5 moments (N=2,4,6,8,10) of each anomalous dimension
$\gamma^{ps}$, $\gamma^{ns}$, $\gamma^{\psi G}$, $\gamma^{G \psi}$ and
$\gamma^{GG}$. 
\begin{equation} \label{ano2loop}  \parbox{17cm}{
$ \begin{array}{lll}
\gamma^{\psi G,(1)}, \gamma^{G\psi,(1)}&:&  \left\{ 1/x,\log(x),1,\log(1-x),
\log^2(1-x)\right\}, \\ & & \left\{1/x,\log^2(x), 1,x,\log^2(1-x)\right\}.\\
 \gamma^{GG,(1)} &:& \left\{1/x,\log^2(x),1,\log^2(1-x),
       \left[\frac{1}{1-x}\right]_{+},\delta(1-x)\right\},\\
& & \left\{ 1/x,\log(x),1,\log(1-x),
  \left[\frac{1}{1-x}\right]_{+},\delta(1-x)\right\}.\\
 \gamma^{ns,(1)} &: & \left\{\log^2(x),\log(x),\log(1-x),
\left[\frac{1}{1-x}\right]_{+},\delta(1-x)\right\}, \\
 & & \left\{ \log^2(x),1,\log^2(1-x),
 \left[\frac{1}{1-x}\right]_{+},\delta(1-x)\right\}\\
\gamma^{ps,(1)} &: & \left\{ 1/x,\log^2(x),\log(x),1,x\right\} \\
 & & \left\{ 1/x,\log^2(x),1,x,x^2 \right\}
\end{array}$ } \vspace{-.6cm}
\end{equation} \\
In these formulae each pair of curly brackets encloses
a set of simple distributions
that appear in a linear combination with coefficients to be matched.
We take 2 different linear combinations for each anomalous dimension
to see the stability of the fit. 

 Please note that all the coefficients in these linear combinations
were taken as arbitrary in the fit (i.e. we even did not fix the
coefficients of the leading singularities) to have a strong
check of the approach.
The only additional information that is used in the reconstruction
is that the coefficients of the most singular large-$x$ terms
$[ 1/(1-x) ]_{+}$ in $\gamma^{\psi\psi}$ and $\gamma^{GG}$
have a simple ratio $C_F/C_A$.

The considered distributions are singular at $x=0$ and $x=1$,
and because e.g. $\delta(1-x)$ can not be drawn
one should consider the figures
only as an indication of the accuracy of the full distributions
which are used in convolutions
(where all the terms that are singular at $x = 1$ fully contribute).
The approximations are stable for $x>.2$
and do dramatically improve in the small-$x$ region
 when more moments (N=12,14,$\cdots$) are used such that
more simple distributions can be included in the linear combinations.
\newpage
\[ \vspace{8.5cm} \]
\[
\parbox{16.5cm}{{\bf Figure 5.} The effects of the NNL order (i.e of the 3-loop 
anomalous dimensions and of the 3-loop beta-function)
on the evolution of the singlet distributions $q^+(x,Q^2)$ and $G(x,Q^2)$.
The parton distributions $q^{+}_{NL}$, $G_{NL}$
and $q^{+}_{NNL}$, $G_{NNL}$ are obtained as solutions of the DGLAP 
evolution equations
(\ref{ap1}) in respectively the NL and NNL approximations starting from the
parametrization scale $Q_0^2 = 4$ GeV$^2$.
The difference between curves for the same $Q^2$ indicates 
the uncertainty in the NNL order effects due to the lack of 
knowledge of the higher anomalous dimensions $\gamma_{N}^{(2)}$, $N > 8$.
The parton distributions at $Q_0^2$ correspond to the MRS(A) 
set \cite{mrsa}. For simplicity, $n_f = 4$
and $\Lambda_{QCD} = 300$ MeV are taken throughout the evolution.
   } \]

The effects of the NNL order on the
evolution of the singlet distributions $q^+(x,Q^2)$ and $G(x,Q^2)$ is
illustrated in figure 5.
The evolutions are done using a program based on the Laguerre polynomial
technique of Ref. \cite{laguerre}.
An estimate of the uncertainty in the NNL effects due to
unknown higher moments is made by fitting various
linear combinations of simple distributions to the calculated
3-loop moments of the anomalous dimensions. The curves correspond to the
sets  of simple distributions of Eq. (\ref{sets3loopano})
(two sets for each anomalous dimension)
which are representative for a larger variation
in the choice of simple distributions.

\[ \parbox{17cm}{
$ \begin{array}{lll}
\gamma^{\psi G,(2)} &:&  \left\{\log(x)/x,\log(x),1,\log(1-x),
             \log^4(1-x)\right\}, \\ 
   & &  \left\{\log(x)/x,1/x,\log(x),\log(1-x),
               \log^3(1-x)\right\}, \\ 
 \gamma^{G\psi,(2)} &:&  \left\{\log(x)/x,\log^2(x),\log^2(1-x),
    \log^4(1-x)\right\}, \\ 
  & &  \left\{1,\log(1-x),\log^2(1-x),\log^3(1-x)\right\},  
\end{array}$ } \vspace{-.6cm} \]
\begin{equation} \parbox{17cm}{
$ \begin{array}{lll}
 \gamma^{GG,(2)} &:& \left\{1,\log^2(1-x),\log^4(1-x),
       \left[\frac{1}{1-x}\right]_{+},\delta(1-x)\right\},\\
& & \left\{ \log(x)/x,1,\log^2(1-x),
  \left[\frac{1}{1-x}\right]_{+},\delta(1-x)\right\}.\\
 \gamma^{ns,(2)} &: & \left\{\log^4(x),1,\log^2(1-x),\log^4(1-x),
\left[\frac{1}{1-x}\right]_{+},\delta(1-x)\right\}, \\
 & & \left\{\log^4(x), \log^2(x),1,\log^2(1-x),
 \left[\frac{1}{1-x}\right]_{+},\delta(1-x)\right\}\\
\gamma^{ps,(2)} &: & \left\{ \log(x)/x,1/x,\log^3(x),1,x^2 \right\} \\
     &  & \left\{ \log(x)/x,\log^4(x),\log^2(x),1,x^2 \right\} \\
\end{array}$ } \vspace{-.6cm}
\label{sets3loopano} \end{equation} \\
The coefficients
of the leading $x\rightarrow 0$ terms of $\gamma^{\psi G,(2)}$,
$\gamma^{ps,(2)}$ and $\gamma^{ns,(2)}$ are known exactly, see Eqs.
 (\ref{leadns},\ref{leads}).
Please notice that
in the variation of simple distributions we changed
even the type of the expected leading singularities as given in
Eq. (\ref{sets3loopano}) (when little information about
these leading singularities is known)
to obtain a conservative estimate of the accuracy of the reconstructions.
One may see that the NNL effects
are of the order of a few percent in the region of $x>0.1$ .
Approximately $50 \%$ of the shown NNL order effect on the
evolution of the parton distributions is due
to the inclusion of the 3-loop beta-function and the other NNL order terms
in the expression for the strong coupling constant, Eq. (\ref{alphaeff}).
The other $50 \%$ is due to the 3-loop anomalous dimensions.
The estimated uncertainty is small in the evolution of
$q^{+}(x)$ and is somewhat larger in the evolution of $G(x)$.
  This is a consequence of the especially good reconstruction of the
    quark anomalous dimensions (see figure 6), which is itself due to
  the existence of separate information about the pure-singlet
  and non-singlet parts.

The curves in figure 6 correspond to matching the following linear combinations
of simple distributions to the calculated moments (where the coefficients
of the leading $x\rightarrow 0$ terms are known exactly, see Eqs.
 (\ref{leadns},\ref{leads}))
\begin{equation} \label{extra3loopano} \parbox{17cm}{
$ \begin{array}{lll}
 \gamma^{ns,(2)} &: & \left\{\log^4(x),\log^2(x),1,\log^2(1-x),
\left[\frac{1}{1-x}\right]_{+},\delta(1-x)\right\}, \\
 & & \left\{\log^4(x),\log(x),\log(1-x),\log^3(1-x),
\left[\frac{1}{1-x}\right]_{+},\delta(1-x)\right\}, \\
 & & \left\{\log^4(x),1,\log^2(1-x),\log^4(1-x),
\left[\frac{1}{1-x}\right]_{+},\delta(1-x)\right\}, \\
\gamma^{ps,(2)} &: & \left\{\log(x)/x, 1/x,\log^4(x),\log^2(x),1 \right\} \\
   &  & \left\{\log(x)/x, 1/x,\log^3(x),1,x^2 \right\} \\
   &  & \left\{\log(x)/x, \log^4(x),\log^2(x),1,x^2 \right\}
\end{array}$ } \vspace{-.6cm}
\end{equation} \\
\newpage
\[ \vspace{8cm} \]   
\[ \parbox{16.5cm}{{\bf Figure 6.} The quality of the reconstruction of the
3-loop quark anomalous dimensions based on the moments
 N=2,4,6,8,(10 for non-singlet only) and the known leading small-$x$ terms
for $\gamma^{ps,(2)}$ and $\gamma^{ns,(2)}$.
The reconstructions are obtained by fitting the sets of
distributions of Eq. (\ref{extra3loopano}) to the available moments. } \]

From the foregoing analysis we have concluded that the inclusion
of the 3-loop anomalous dimensions
in the evolution of parton distributions in the $\overline{\rm MS}$-scheme
changes both the quark and
gluon distributions by 1-3 $\%$ in the region $0.1<x<0.8 $ for a 
change in $Q^2$ between $(2$ GeV$)^2$ and $(100$ GeV$)^2$.

We further remark that it is important to obtain the
exact results for the 3-loop anomalous dimensions for all $N$ to do a
complete analysis in the NNL order.

\subsection{Coefficient Functions}
As an example of the effects of the calculated 3-loop coefficient functions
on the deep inelastic structure functions we consider the case of $C_2^{ns}$. 
We use the following notation for the inverse Mellin transformed 
coefficient function.
\begin{eqnarray} C_2^{ns}(x,a_s)& = & C_2^{ns,(0)}(x) + a_s C_2^{ns,(1)}(x)
    + a_s^2 C_2^{ns,(2)}(x)+ a_s^3 C_2^{ns,(3)}(x) + O(a_s^4). 
 \end{eqnarray}
Please note that for the complete NNNL order
approximation for $F_2$ one would need, besides the 3-loop $C_2^{ns,(3)}$,
 the 4-loop anomalous dimensions which are difficult to calculate at present.

Presuming that the anomalous dimensions 
are not more singular than $\log( N) $ as $ N\rightarrow \infty$
one has shown that all the logarithms $\log^i ( N)$ that are present
in the non-singlet parts of the coefficient functions exponentiate
according to the soft gluon resummation formulae 
\cite{softpar,softster,softcat,softlaen}.
We will use this to obtain the large N limit of moments of the 
3-loop $C^{ns}_{2}$ coefficient function 
(i.e. the $x\rightarrow 1$ limit of $C^{ns}_{2}$).
Exponentiating the $\log^i( N) $ terms from the explicitly known 1 and 2-loop
coefficient functions we find
 \begin{eqnarray}
 C^{ns}_{2,N}  & \stackrel{N \rightarrow \infty}{ \approx } 
 \exp \Biggl\{  & a_s \left[ 2C_F \log^2 (N)+C_F (3+4\gamma_E) \log(N) \right] 
+ a_s^2 \Bigl[  \frac{2}{3} C_F (\frac{11}{3} C_A-\frac{2}{3} n_f)\log^3 (N) 
  \nonumber \\ & & + O( \log^2(N) ) \Bigr] 
   + O( a_s^3 \log^4 (N) ) \Biggr\} + O(a_s N^0) 
 \end{eqnarray}  

\noindent 
where $\gamma_E$ is the Euler constant.
In $x$ space this corresponds to\footnote{for a table of the moments 
 of the relevant distributions in the large $N$ limit,
see e.g. Ref \cite{softcat}.}  \\
\begin{equation}
\parbox{17cm}{
$ \begin{array}{lll}
&  C^{ns}_{2}(x)   \stackrel{x\rightarrow 1}{ \approx }
 & 1 + a_s  \Biggl\{   4 C_F \left[ \frac{ \log (1-x) }{1-x}\right]_{+}
        - 3 C_F \left[ \frac{ 1 }{1-x}\right]_{+}
        + O( \delta (1-x)  ) \Biggr\} 
 + a_s^2  \Biggl\{  8 C_F^2  \left[ \frac{ \log^3 (1-x) }{1-x}\right]_{+} \\
& &                     + (  - \frac{22}{3} C_A C_F + \frac{4}{3}C_F n_f
                - 18 C_F^2 )  \left[ \frac{ \log^2 (1-x) }{1-x}\right]_{+}
 + O\left( \left[ \frac{ \log (1-x) }{1-x}\right]_{+} \right) \Biggr\} \\
& & + a_s^3  \Biggl\{  8 C_F^3 \left[ \frac{ \log^5 (1-x) }{1-x}\right]_{+}
      + ( -\frac{220}{9}C_A C_F^2+ \frac{40}{9} C_F^2 n_f-30 C_F^3 )
                \left[ \frac{ \log^4 (1-x) }{1-x}\right]_{+} 
 \\
& & + O\left( \left[ \frac{ \log^3 (1-x) }{1-x}\right]_{+} \right)  \Biggr\}
    + O\left( a_s^4 
        \left[ \frac{ \log^7 (1-x) }{1-x}\right]_{+} \right). 
 \\ \end{array} $} \vspace{-.8cm}
\label{softx} 
\end{equation} \\

The non-singlet coefficient function $C^{ns}$ does not have $1/x$ terms
for small-$x$ (it is finite as $N \rightarrow 1$)
and the leading small-$x$ term that one may expect at 3-loops 
(from extrapolating the 1-loop and 2-loop results) is $\log^{5}(x)$.
(The singlet coefficient function $C^{\psi}_2$ contains the same large-$x$ terms
since it contains the same non-singlet contribution.
The small-$x$ limits of the singlet coefficient functions do contain
terms of the type $\log^i (x)/x$, $i \geq 0$.
 At 3-loops one expects a term $\log(x)/x$.)  

At the two loop order the reconstruction of the coefficient functions
can be compared with the exact results. In figure 7 we illustrated
the quality of the reconstruction based on matching the following sets 
of simple distributions to the first 5 moments N=2,4,6,8,10  

\begin{equation} \label{2loopC2sets} \parbox{17cm}{
$ \begin{array}{lll}
 C_2^{ns,(2)} &:&
\left\{ \log(x),1,x^2,\delta(1-x),
\left[\frac{1}{1-x}\right]_{+},
\left[\frac{\log^2(1-x)}{1-x}\right]_{+},
\left[\frac{\log^3(1-x)}{1-x}\right]_{+}\right\} \\
& &
\left\{ \log^2(x),1,\delta(1-x),
\left[\frac{1}{1-x}\right]_{+},
\left[\frac{\log(1-x)}{1-x}\right]_{+},
\left[\frac{\log^3(1-x)}{1-x}\right]_{+}\right\}
\end{array} $ } \vspace{-.6cm}
\end{equation} \\
It is understood that for the first set the coefficients of the last
two terms  are taken from Eq. (\ref{softx}), and that for the second set 
only the coefficient of $[ \log^3(1-x)/(1-x) ]_{+}$ was taken exactly. 
One can see from
figure 7 that the reconstruction of the distribution $C_2^{ns,(2)}$
is quite stable for $x > 0.2$ .
We want to emphasize that considering {\em only} the two most singular
soft gluon terms (i.e.
$[ \log^3(1-x)/(1-x) ]_{+}$ and $[ \log^2(1-x)/(1-x) ]_{+}$)
gives a huge overestimate of the true effects.
However, as it was observed in Ref. \cite{nz},
all the terms of the type $[ \log^i(1-x)/(1-x) ]_{+}$, $i \geq 0$, together
constitute a large fraction of $C_2^{ns,(2)} \otimes \Delta$ in the region
$x > 0.5$ (more precisely we obtain 40 \% at  $x = 0.6$ and  
95 \% at $x = 0.75$).

\[ \vspace{8cm} \]
\[ \parbox{16.5cm}{{\bf Figure 7.} The exact 
2-loop coefficient function $C_2^{ns,(2)}$
and reconstructions (approxs.)
based on the lowest 5 moments N=2,4,6,8,10 and the sets of distributions
of Eq. (\ref{2loopC2sets}).
The left plot shows the reconstruction for intermediate $x$. The
right plot shows the quality of the reconstructed 
coefficient function as a distribution in the whole $x$-region, 
i.e. when it is convoluted with
a realistic parton distribution $\Delta = u+\overline{u}-d-\overline{d}$.
The quark distributions correspond to the MRS(A) set \cite{mrsa}
at its parametrization scale $Q_0^2 = 4$ GeV$^2$, $n_f=4$.} \]
The present status of the effects of the higher order contributions 
to $C^{ns}_2$ is illustrated in figure 8. The 3-loop contribution is
obtained by matching the following 4 linear combinations of distributions 
\begin{equation} \label{sets3loopcoef} \parbox{17cm}{
$ \begin{array}{lll}
 C_2^{ns,(3)} &:& \left\{1,x^2,\delta(1-x),\left[\frac{1}{1-x}\right]_{+},
\left[\frac{\log^2(1-x)}{1-x}\right]_{+},
\left[\frac{\log^4(1-x)}{1-x}\right]_{+},
\left[\frac{\log^5(1-x)}{1-x}\right]_{+}\right\} \\
& &
\left\{ \log(x),1,x^2,\delta(1-x),
\left[\frac{\log^2(1-x)}{1-x}\right]_{+},
\left[\frac{\log^4(1-x)}{1-x}\right]_{+},
\left[\frac{\log^5(1-x)}{1-x}\right]_{+}\right\} \\
& &
\left\{ \log(x),1,x^2,\delta(1-x),
\left[\frac{1}{1-x}\right]_{+},
\left[\frac{\log^4(1-x)}{1-x}\right]_{+},
\left[\frac{\log^5(1-x)}{1-x}\right]_{+}\right\} \\
& &
\left\{ \log^2(x),1,x,\delta(1-x),
\left[\frac{1}{1-x}\right]_{+},
\left[\frac{\log^4(1-x)}{1-x}\right]_{+},
\left[\frac{\log^5(1-x)}{1-x}\right]_{+}\right\} 
\end{array} $  } \vspace{-.6cm}
\end{equation} \\
to the  5 available  moments (N=2,4,6,8,10),
where the numerical coefficients of the terms $[\log^5(1-x)/(1-x)]_{+}$
and $[\log^4(1-x)/(1-x)]_{+}$ are taken from Eq. (\ref{softx}).
Please notice that
in the variation of simple distributions in the 4 linear combinations
above we changed even the type of the expected small-$x$ singularities 
to obtain a conservative estimate of the accuracy of the reconstructions.

 \[
\parbox{14cm}{
\vspace{15.5cm} 
 \parbox{14cm}{{\bf Figure 8.}
Higher order corrections to $F_2$ due to the non-singlet coefficient function 
$C^{ns}_2$.
The different curves show the effects on $F_2$ of higher order corrections to
$C^{ns}_2$ via the relation
 $\frac{1}{x}F_2^{ep}(x,Q^2)-\frac{1}{x}F_2^{en}(x,Q^2) =
   C^{ns}_2 \otimes \frac{1}{3}( u+\overline{u}-d-\overline{d})$,
when both the quark distributions and $\alpha_s$ are kept constant.
The spread of the different NNNLO curves indicates the uncertainty of the
3-loop contribution (due to variations of the distributions as it is
given in Eq. (\ref{sets3loopcoef})).
We used the parton distributions in the $\overline{\rm MS}$ scheme
from the MRS(A) global fit \cite{mrsa} (in the NL order)
at its parametrization scale $Q^2 = 4$ GeV$^2$ and $\alpha_s = .261$, $n_f=4$. 
(LO = leading order, NLO = next-to-leading order, etc.)

 } }
\]

For the non-singlet combination of the nucleon structure functions
$F_2^{ep}(x,Q^2)-F_2^{en}(x,Q^2) $
that is presented in figure 8, $C_2^{ns}$ is the only relevant
coefficient function.
The different curves show the effects on $F_2$ of higher order corrections to
$C^{ns}_2$ when both the quark distributions and $\alpha_s$ are kept constant.
Although the change in $F_2$ is indicated when the order of
approximation for $C_2$ changes
(LO indicates $C_2^{ns,(0)}$, 
NLO indicates $C_2^{ns,(0)}+a_s C_2^{ns,(1)}$, etc.),
it is in fact the quark distributions that may have to be modified to keep
$F_2$ in agreement with the experiments over a large $(x,Q^2)$ range
when higher order corrections are globally included.
We noticed that also at 3-loops the two leading soft gluon contributions
are to a large extent suppressed by `subleading' contributions in the
coefficient functions.

At a relatively low energy scale of $Q^2= 4$ GeV$^2$ of figure 8
we find that the 3-loop contribution $a_s^3 C_2^{ns,(3)}$
gives a sizable correction to the coefficient function $C_2^{ns}$
(about 1/3 of the 2-loop contribution $a_s^2 C_2^{ns,(2)}$
in the $x$ region $0.2 \leq x \leq 0.9$).
Furthermore, from figure 8 one can observe apparent convergence
of the QCD perturbation theory up to and including the NNNL order.

\section{The analytic results of the 3-loop calculation}

Before we present the results for the anomalous dimensions and coefficient
functions we should explain our conventions for the different
flavour factors that appear in the present 3-loop calculation.  \\
\[ \vspace{1.4cm}\]
\[  FC_2 \hspace{1.8cm} FC_{11} \hspace{1.8cm} FC_{02}
 \hspace{1.8cm} FC^{g}_{2}  \hspace{2.0cm}FC^{g}_{11} \] 
\[ \parbox{15cm}{ {\bf Figure 9.}
Examples for diagrams in the flavour classes $FC_2$, $FC_{11}$, $FC_{02}$,
 $FC^{g}_{2}$, $FC^{g}_{11}$.
} \] \vspace{0cm}

Diagrams in the flavour class $FC_{2}$ (where both photons are attached to
 the quark line of the external quark legs) have a SU$(n_f)$ flavour 
factor  $\hat{Q}_f^2$ where the matrix $\hat{Q}_f $
 is the quark charge matrix from the electromagnetic current 
($\hat{Q}_f = $diag($\frac{2}{3}$,$-\frac{1}{3}$,$-\frac{1}{3}$,$\cdots$)
and tr$( \hat{Q_f}) = \sum_{f=1}^{n_f} e_f $ where
$e_f$ is the electromagnetic charge of a quark with the corresponding flavour).
Diagrams in the flavour
class $FC_{11}$ (where one photon is  attached to a closed quark loop and
the other one to the quark line of the external quark legs)
have a flavour matrix  tr$\left(\hat{Q}_f\right) \hat{Q}_f $
and diagrams in the flavour class $FC_{02}$ (flavour singlet diagrams, 
where both photon legs are  attached to the same internal quark loop)  
have a flavour matrix tr$\left(\hat{Q}_f^2\right) {\large \bf 1} $.
The diagrams with external gluons are split up into two flavour classes:
the class $FC^{g}_{2}$ has a flavour factor  tr$(\hat{Q}_f^2) $ and
the class $FC^{g}_{11}$ has a flavour factor  (tr$(\hat{Q}_f))^2 $.

To project out the non-singlet contributions one should take the 
flavour trace of the diagrams with the generators $\lambda^{\alpha}$
of SU$(n_f)$ (due to the diagonal form of $\hat{Q_f}$ only the
diagonal generators are relevant). This means that in the non-singlet
projection only the diagrams from the flavour classes $FC_{2}$ and $FC_{11}$
contribute.
In this way the diagrams of the class $FC_{2}$
receive a flavour factor tr$(\hat{Q_f}^2\lambda^{\alpha})$,
and the diagrams of the class $FC_{11}$ receive a
flavour factor tr$(\hat{Q_f})$tr$(\hat{Q_f}\lambda^{\alpha})$.
Since the ratio of these flavour factors does not depend on the number
 $\alpha$ of a diagonal generator $\lambda^{\alpha}$
\begin{equation}
 \frac{{\rm tr}(\hat{Q_f}\lambda^{\alpha})}{{\rm tr}(\hat{Q_f}^2
\lambda^{\alpha})} =3  \label{factor3} \end{equation}
it is possible to factorize the $\alpha$-independent coefficient 
functions $C_{k,N}^{ns}$ as follows
\[ \sum_{\alpha} C_{k,N}^{\alpha}A_{{\rm nucl},N}^{\alpha} =  
\left[ C_{k,N}(FC_2)+
\frac{{\rm tr}(\hat{Q_f}){\rm tr}(\hat{Q_f}\lambda^{\beta})}
{{\rm tr}(\hat{Q_f}^2\lambda^{\beta})}C_{k,N}(FC_{11})\right]\times
\left[ \left(\sum_{\alpha}{\rm tr}(\hat{Q_f}^2\lambda^{\alpha}\right)
 A_{{\rm nucl},N}^{\alpha} \right] \]
\begin{equation}  \label{factor}
 =  C_{k,N}^{ns}\left(\frac{Q^2}{\mu^2},a_s\right)\times 
   A_{{\rm nucl},N}^{ns} (p^2/\mu^2), 
     \hspace{5 cm}
\end{equation} 
where the non-singlet $C_{k,N}^{ns}$ is independent of $\alpha$ and the
standard normalization is $C_{k,N}^{ns} = 1 + O(a_s)$.

To project out the flavour singlet contributions for diagrams with external
quark legs one should take the trace with the unit flavour matrix.

The explicit values for the flavour factors  $fl_2$, $fl_{11}$, $fl_{02}$,
 $fl^g_2$, $fl^g_{11}$ (for the flavour classes
 $FC_2$, $FC_{11}$, $FC_{02}$,
 $FC^{g}_{2}$ and $FC^{g}_{11}$ respectively)
that give the standard normalization of the coefficient functions  
for the non-singlet and singlet cases are given in table 2.
Please notice that singlet flavour factors are chosen such that they 
reduce to unity if one substitutes for $\hat{Q_f}$ the unit matrix 
$\hat{Q_f} = diag(1,1,1,\cdots)$. Please notice that the factor 3 in the
non-singlet flavour factor $fl_{11}$ originates from Eq. (\ref{factor3}).
{ \[ \begin{tabular}{l c c c c c}\\
\hline
{} & $fl_2$ & $fl_{11}$ & $fl_{02}$ & $fl^{g}_{2}$ & $fl^{g}_{11}$ \\
\hline
non-singlet & 1  & $ \frac{3}{nf} \sum_{f=1}^{n_f} e_f $  & 0 & --- & --- \\ \\
singlet     & 1  & $ \frac{1}{n_f} \frac{ \left(\sum_{f=1}^{n_f} e_f\right)^2}{
      \sum_{f=1}^{n_f} e_f^2 } $  & 1 & 1 &
 $ \frac{1}{n_f} \frac{ \left(\sum_{f=1}^{n_f} e_f\right)^2 }{
      \sum_{f=1}^{n_f} e_f^2 }$ \\
\hline\\
\end{tabular} \] 
\[ \parbox{14cm}{ {\bf Table 2.}
Values for the flavour factors $fl_2$, $fl_{11}$, $fl_{02}$,
 $fl^g_2$, $fl^g_{11}$.
} \] }

Below we present the results for the 3-loop anomalous dimensions
and the 3-loop coefficient functions.
Please notice that the results contain both the singlet 
and non-singlet contributions. To obtain the specific non-singlet or
singlet case one should substitute the flavour factors from
table 2. In the results we have already put $fl_2 = fl^g_2 = 1$.
The reader may notice an overall factor 2 difference between our results
for anomalous dimensions
and the results in Ref. \cite{frs} which originates from our definition
of the renormalization scale derivative $d/(d \ln (\mu^2))$
 (see Eqs. (\ref{defgammas}) and (\ref{defgammans}))
versus the derivative $d/(d \ln ( \mu ))$ in  Ref. \cite{frs}.
In the results for the coefficient functions we have put $\mu^2 = Q^2$ 
(in accordance with the evolution equation Eq.(\ref{rg})).
The Riemann zeta function is written as $\zeta_n$.
\[ \]
\renewcommand{\arraystretch}{ 1.2}
\noindent $ \begin{array}{lll}
 \gamma_2^{\psi\psi}  &  = & \hspace{3mm} a_s  C_F \frac{8}{3}  \\ 
& &  + a_s^2  \Bigl[ 
   C_F C_A \frac{376}{27} + C_F^2 (- \frac{112}{27})
   +  n_f C_F ( -\frac{64}{27})  \Bigr] \\
& & +  a_s^2  fl_{02} C_F n_f (-\frac{40}{27})  \\
& &  + a_s^3  \Bigl[ 
      C_F C_A^2  ( \frac{20920}{243} + \frac{64}{3} \zeta_3 ) 
   +  C_F^2 C_A  (  - \frac{8528}{243} - 64 \zeta_3 )
   +  C_F^3  (  - \frac{560}{243} + \frac{128}{3} \zeta_3 ) \\
& & \hspace{5mm}
   +  n_f C_F C_A  (  - \frac{3128}{243} - \frac{64}{3} \zeta_3 )
   +   n_f C_F^2  (  - \frac{3412}{243} + \frac{64}{3} \zeta_3 )
   +   n_f^2 C_F  (  - \frac{224}{243} )    \Bigr] \\ 
& &  + a_s^3  fl_{02} \Bigl[
    C_F C_A n_f  ( \frac{2534}{243} - \frac{64}{3} \zeta_3 )
  +  C_F^2 n_f  (  - \frac{3682}{243} + \frac{64}{3} \zeta_3 )
  +  C_F n_f^2  (  - \frac{628}{243} ) \Bigr]  \\
& = & \hspace{3mm} a_s  3.555555556  \\
& &+ a_s^2  \Bigl( 48.3292181070 -3.1604938272 n_f \Bigr) \\
& &+ a_s^2  fl_{02} \Bigl(  -1.9753086420 n_f \Bigr) \\
& &+ a_s^3  \Bigl(  859.4478371772  - 133.4381617374  n_f
           -1.2290809328  n_f^2 \Bigr)  \\
& &+ a_s^3  fl_{02} \Bigl( 
  -42.2118242888 n_f  -3.4458161866 n_f^2 \Bigr) \\
\end{array} $ \\
\vspace{.3cm}

\noindent $ \begin{array}{lll}
 \gamma_4^{\psi\psi}  &  = & \hspace{3mm} a_s C_F \frac{157}{30}  \\
& &  + a_s^2 \Bigl[
  C_F C_A  \frac{16157}{675}  +C_F^2 ( -\frac{287303}{54000} ) 
  + n_f C_F ( -\frac{13271}{2700} )   \Bigr] \\
& & +  a_s^2 fl_{02}  n_f C_F (-\frac{2147}{27000} ) \\
& &  + a_s^3 \Bigl[
   C_F C_A^2  ( \frac{136066373}{972000} + \frac{1439}{75} \zeta_3 )
  + C_F^2 C_A  (  - \frac{267028157}{9720000} - \frac{1439}{25} \zeta_3 ) \\
& & \hspace{5mm}
   + C_F^3  (  - \frac{714245693}{48600000} + \frac{2878}{75} \zeta_3 ) 
   + n_f C_F C_A (  - \frac{8802581}{486000} - \frac{628}{15} \zeta_3 ) \\
& & \hspace{5mm}
   + n_f C_F^2  (  - \frac{165237563}{4860000} + \frac{628}{15} \zeta_3 )
   + n_f^2 C_F  (  - \frac{384277}{243000} ) \Bigr]  \\
& &  + a_s^3 fl_{02} \Bigl[
    n_f C_F C_A  ( \frac{2485097}{2025000} - \frac{242}{75} \zeta_3 )
   + n_f C_F^2  (  - \frac{2217031}{2700000} + \frac{242}{75} \zeta_3 )
   + n_f^2 C_F  (  - \frac{618673}{1215000} ) \Bigr]  \\
& = & \hspace{3mm} a_s 6.9777777778  \\
& &+ a_s^2  \Bigl( 86.2866502058  -6.5535802469 n_f \Bigr) \\
& &+ a_s^2  fl_{02} \Bigl(  -0.1060246914 n_f \Bigr) \\
& &+ a_s^3  \Bigl( 1515.5623634355 -244.7285919523 n_f
        -2.1085157750 n_f^2 \Bigr) \\
& &+ a_s^3  fl_{02} \Bigl( -5.1701331196 n_f
     -0.6789278464 n_f^2 \Bigr) 
\end{array} $ \\
\vspace{.3cm}

\noindent $ \begin{array}{lll}
 \gamma_6^{\psi\psi}  &  = & \hspace{3mm} a_s C_F \frac{709}{105}  \\
& &  + a_s^2 \Bigl[
  C_F C_A  \frac{157415}{5292}  +C_F^2 (  - \frac{3173311}{514500} )
  + n_f C_F (  - \frac{428119}{66150} )   \Bigr] \\
& & +  a_s^2 fl_{02}  n_f C_F (  - \frac{289}{18522} ) \\
& &  + a_s^3 \Bigl[
   C_F C_A^2  (  \frac{115237918583}{666792000} + \frac{69862}{3675} \zeta_3 )
  + C_F^2 C_A  ( - \frac{34855421369}{1166886000}
   - \frac{69862}{1225} \zeta_3  ) \\
& & \hspace{5mm}
   + C_F^3  (  - \frac{854652999073}{51051262500} + \frac{139724}{3675}\zeta_3 )
   + n_f C_F C_A ( - \frac{13978373}{686000} - \frac{5672}{105} \zeta_3  ) \\
& & \hspace{5mm}
   + n_f C_F^2  (  - \frac{44644018231}{972405000} + \frac{5672}{105} \zeta_3 )
   + n_f^2 C_F  (  - \frac{80347571}{41674500}  ) \Bigr]  \\
& &  + a_s^3 fl_{02} \Bigl[
    n_f C_F C_A  (  \frac{1988624681}{16336404000} - \frac{968}{735} \zeta_3 )
   + n_f C_F^2  (  \frac{11602048711}{40841010000} + \frac{968}{735} \zeta_3  )
   + n_f^2 C_F  (   - \frac{31555763}{145860750}  ) \Bigr]  \\
& = & \hspace{3mm} a_s  9.0031746032  \\
& &+ a_s^2  \Bigl( 108.0184697117 - 8.6292567397 n_f \Bigr) \\
& &+ a_s^2  fl_{02} \Bigl( - 0.0208040888 n_f
          \Bigr) \\
& &+ a_s^3  \Bigl( 1891.8277788314 -307.4236889402 n_f
        -2.5706389919 n_f^2  \Bigr) \\
& &+ a_s^3  fl_{02} \Bigl( -2.5260911925 n_f
    -0.2884556035 n_f^2  \Bigr)
\end{array} $  \\
\vspace{.3cm}

\noindent $ \begin{array}{lll}
 \gamma_8^{\psi\psi}  &  = & \hspace{3mm} a_s C_F \frac{9883}{1260}  \\
& &  + a_s^2 \Bigl[
  C_F C_A  \frac{25870049}{762048} +C_F^2  ( - \frac{27040578211}{4000752000} )
  + n_f C_F (  - \frac{36241943}{4762800} )   \Bigr] \\
& & +  a_s^2 fl_{02}  n_f C_F (  - \frac{40333}{8164800} ) \\
& &  + a_s^3 \Bigl[
   C_F C_A^2 ( \frac{8101059985033}{41150592000} 
                  + \frac{2510407}{132300}\zeta_3 ) 
  + C_F^2 C_A (  - \frac{3662576699059}{112021056000} - 
                  \frac{2510407}{44100}\zeta_3 )
    \\
& & \hspace{5mm}
   + C_F^3  (  - \frac{109308710097437993}{6351593875200000} 
   + \frac{2510407}{66150}\zeta_3 )
   + n_f C_F C_A   (  - \frac{1578915745223}{72013536000}
         - \frac{19766}{315} \zeta_3 ) \\
& & \hspace{5mm}
   + n_f C_F^2 (  - \frac{91675209372043}{1680315840000} 
            + \frac{19766}{315} \zeta_3 )
   + n_f^2 C_F (  - \frac{38920977797}{18003384000} )  \Bigr]  \\
& &  + a_s^3 fl_{02} \Bigl[
    n_f C_F C_A  (  - \frac{343248329803}{2592487296000}
         - \frac{1369}{1890}\zeta_3 )
   + n_f C_F^2  ( \frac{39929737384469}{90737055360000} 
                     + \frac{1369}{1890} \zeta_3 ) \\
& & \hspace{5mm} 
 + n_f^2 C_F  (  - \frac{13131081443}{108020304000} )  \Bigr]  \\
& = & \hspace{3mm} a_s 10.4582010582   \\
& &+ a_s^2  \Bigl( 123.7764525165  - 10.1458366227 n_f \Bigr) \\
& &+ a_s^2  fl_{02} \Bigl( - 0.0065864851 n_f \Bigr) \\
& &+ a_s^3  \Bigl(  2164.0918358230 - 352.3116595904 n_f
       - 2.8824934836  n_f^2   \Bigr) \\
& &+ a_s^3  fl_{02} \Bigl( - 1.6821565188 n_f
     - 0.1620816452 n_f^2  \Bigr)
\end{array} $ \\
\vspace{.3cm}

\noindent $ \begin{array}{lll}
 \gamma_2^{\psi G}  &  = & \hspace{3mm} a_s n_f (- \frac{2}{3} )  \\
& &  + a_s^2 \Bigl[ n_f C_F (-\frac{74}{27}) + n_f C_A (-\frac{35}{27})
      \Bigr] \\
& &  + a_s^3 \Bigl[  n_f C_F C_A ( \frac{139}{9} - \frac{104}{3} \zeta_3 )
       + n_f C_F^2  (  - \frac{2155}{243} + \frac{32}{3} \zeta_3 ) \\
& & \hspace{5mm} 
      + n_f C_A^2  (  - \frac{3589}{162} + 24 \zeta_3 )
      + n_f^2 C_F  (  - \frac{173}{243} )
      + n_f^2 C_A  ( \frac{1058}{243} ) \Bigr] \\
& = & \hspace{3mm} a_s \Bigl( - 0.6666666667 n_f \Bigr) \\
& &+ a_s^2 \Bigl( - 7.5432098765 n_f \Bigr) \\
& &+ a_s^3 \Bigl( -37.6233727456 n_f+12.1124828532 n_f^2 \Bigr) 
\end{array} $  \\
\vspace{.3cm}

\noindent $ \begin{array}{lll}
 \gamma_4^{\psi G}  &  = & \hspace{3mm} a_s n_f ( -\frac{11}{30} ) \\
& &  + a_s^2 \Bigl[ n_f C_F (  - \frac{56317}{18000} ) 
      + n_f C_A ( \frac{16387}{9000} ) \Bigr] \\
& &  + a_s^3 \Bigl[  n_f C_F C_A ( \frac{278546497}{8100000}
         - \frac{969}{25}\zeta_3 ) 
  + n_f C_F^2  (  - \frac{757117001}{48600000} + \frac{261}{25}\zeta_3 )\\
& & \hspace{5mm}
   + n_f C_A^2 (  - \frac{295110931}{12150000} + \frac{708}{25}\zeta_3 ) 
   + n_f^2 C_F ( \frac{9613841}{24300000} )
   + n_f^2 C_A ( \frac{4481539}{2430000} )   \Bigr] \\
& = & \hspace{3mm} a_s \Bigl(  - 0.3666666667 n_f  \Bigr) \\
& &+ a_s^2 \Bigl(  1.2907037037 n_f      \Bigr) \\
& &+ a_s^3 \Bigl( 33.5814927342 n_f
      + 6.0602726200 n_f^2    \Bigr)
\end{array} $ \\
\vspace{.3cm}

\noindent $ \begin{array}{lll}
 \gamma_6^{\psi G}  &  = & \hspace{3mm} a_s n_f ( -\frac{11}{42} ) \\
& &  + a_s^2 \Bigl[ n_f C_F  (  - \frac{296083}{92610} )
      + n_f C_A ( \frac{867311}{370440} )  \Bigr] \\
& &  + a_s^3 \Bigl[  n_f C_F C_A ( \frac{1199181909343}{32672808000}
                       - \frac{23666}{735} \zeta_3 )
       + n_f C_F^2 (  - \frac{2933980223981}{163364040000} 
                       + \frac{1240}{147} \zeta_3 ) \\
& & \hspace{5mm}  + n_f C_A^2 (  - \frac{58595443051}{2613824640} 
                      + \frac{5822}{245} \zeta_3 ) 
 + n_f^2 C_F ( \frac{1539874183}{2722734000} )
      + n_f^2 C_A  ( \frac{86617163}{93350880} )  \Bigr] \\
& = & \hspace{3mm} a_s \Bigl(  - 0.2619047619 n_f   \Bigr) \\
& &+ a_s^2 \Bigl( 2.7611048123 n_f  \Bigr) \\
& &+ a_s^3 \Bigl(  33.4160213457 n_f
     +  3.5376821018 n_f^2   \Bigr)
\end{array} $  \\
\vspace{.3cm}

\noindent $ \begin{array}{lll}
 \gamma_8^{\psi G}  &  = & \hspace{3mm} a_s n_f (- \frac{37}{180} ) \\
& &  + a_s^2 \Bigl[ n_f C_F  (  - \frac{51090517}{16329600} )
      + n_f C_A  ( \frac{100911011}{40824000} )   \Bigr] \\
& &  + a_s^3 \Bigl[  n_f C_F C_A ( \frac{4896295442015177}{129624364800000}
                   - \frac{515201}{18900} \zeta_3 )
       + n_f C_F^2 (  - \frac{4374484944665803}{226842638400000}
                   + \frac{749}{108} \zeta_3 ) \\
& & \hspace{5mm}  + n_f C_A^2 
               (  - \frac{24648658224523}{1157360400000} 
                   + \frac{64021}{3150} \zeta_3 )
 + n_f^2 C_F ( \frac{7903297846481}{12962436480000} )
      + n_f^2 C_A ( \frac{10379424541}{22044960000} )   \Bigr] \\
& = & \hspace{3mm} a_s \Bigl( - 0.2055555556 n_f \Bigr) \\
& &+ a_s^2 \Bigl(  3.2439572229 n_f  \Bigr) \\
& &+ a_s^3 \Bigl(  28.7612614990 n_f+2.2254331118 n_f^2 \Bigr)
\end{array} $  \\

\newpage
\[ \vspace{-1.8cm} \]
\noindent  $ \begin{array}{lll}
 \gamma_2^{G \psi}  &  = & \hspace{3mm} a_s C_F (- \frac{8}{3} )  \\
& &  + a_s^2 \Bigl[
  C_F C_A (  - \frac{376}{27} )   +C_F^2 ( \frac{112}{27} )
  + n_f C_F ( \frac{104}{27} )   \Bigr] \\
& &  + a_s^3 \Bigl[
   C_F C_A^2 (  - \frac{20920}{243} - \frac{64}{3} \zeta_3 )
  + C_F^2 C_A ( \frac{8528}{243} + 64 \zeta_3 )  \\
& & \hspace{5mm}
   + C_F^3  ( \frac{560}{243} - \frac{128}{3} \zeta_3 )
   + n_f C_F C_A ( \frac{22}{9} + \frac{128}{3} \zeta_3 ) \\
& & \hspace{5mm}
   + n_f C_F^2 ( \frac{7094}{243} - \frac{128}{3} \zeta_3 ) 
   + n_f^2 C_F ( \frac{284}{81} ) \Bigr] \\
& = & \hspace{3mm} a_s \Bigl( -3.5555555556 \Bigr)  \\
& &+ a_s^2  \Bigl( - 48.3292181070 + 5.1358024691  n_f \Bigr) \\
& &+ a_s^3  \Bigl( -859.4478371772 +175.6499860261 n_f
   +4.6748971193 n_f^2 \Bigr) \\
\end{array} $ \\
\vspace{.3cm}

\noindent $ \begin{array}{lll}
 \gamma_4^{G \psi}  &  = & \hspace{3mm} a_s C_F ( -\frac{11}{15} )  \\
& &  + a_s^2 \Bigl[
  C_F C_A  (  - \frac{63949}{13500} )  +C_F^2 ( \frac{42109}{27000} )
  + n_f C_F ( \frac{313}{675} )   \Bigr] \\
& &  + a_s^3 \Bigl[
   C_F C_A^2 (  - \frac{325575847}{12150000} - \frac{944}{75} \zeta_3 )
  + C_F^2 C_A (  - \frac{117100723}{24300000} + \frac{944}{25} \zeta_3 )  \\
& & \hspace{5mm}
   + C_F^3  ( \frac{110687611}{6075000} - \frac{1888}{75} \zeta_3 )
   + n_f C_F C_A  (  - \frac{799}{48600} + \frac{176}{15} \zeta_3 ) \\
& & \hspace{5mm}
   + n_f C_F^2 ( \frac{3651671}{759375} - \frac{176}{15} \zeta_3 )
   + n_f^2 C_F ( \frac{54731}{40500} )  \Bigr] \\
& = & \hspace{3mm} a_s \Bigl( -0.9777777778 \Bigr)  \\
& &+ a_s^2  \Bigl(  - 16.1752427984 + 0.6182716049 n_f \Bigr) \\
& &+ a_s^3  \Bigl( -315.2762549542 +39.8257102724 n_f
 +1.8018436214 n_f^2 \Bigr) \\
\end{array} $ \\
\vspace{.3cm}

\noindent $ \begin{array}{lll}
 \gamma_6^{G \psi}  &  = & \hspace{3mm} a_s C_F ( - \frac{44}{105} ) \\
& &  + a_s^2 \Bigl[
  C_F C_A (  - \frac{437033}{154350} )   +C_F^2 ( \frac{1191271}{1157625} )
  + n_f C_F ( \frac{2204}{33075} )   \Bigr] \\
& &  + a_s^3 \Bigl[
   C_F C_A^2   (  - \frac{202913519537}{16336404000}
                       - \frac{37592}{3675}\zeta_3 )
  + C_F^2 C_A (  - \frac{150561431231}{10210252500}
                   + \frac{37592}{1225} \zeta_3 )  \\
& & \hspace{5mm}
   + C_F^3 ( \frac{1031558954593}{51051262500} - \frac{75184}{3675}\zeta_3 )
   + n_f C_F C_A (  - \frac{253841107}{583443000} + \frac{704}{105} \zeta_3 ) \\
& & \hspace{5mm}
   + n_f C_F^2 ( \frac{20157323311}{10210252500} - \frac{704}{105} \zeta_3 )
   + n_f^2 C_F ( \frac{2833459}{3472875} )  \Bigr] \\
& = & \hspace{3mm} a_s  \Bigl( -0.5587301587 \Bigr) \\
& &+ a_s^2  \Bigl(  - 9.4963177963 + 0.0888485765 n_f \Bigr) \\
& &+ a_s^3  \Bigl( -188.9088124238 +19.6794454593 n_f
  +1.0878437414 n_f^2 \Bigr) \\
\end{array} $ \\
\vspace{.3cm}

\noindent $ \begin{array}{lll}
 \gamma_8^{G \psi}  &  = & \hspace{3mm} a_s C_F ( -\frac{37}{126} )  \\
& &  + a_s^2 \Bigl[
  C_F C_A (  - \frac{58805263}{28576800} ) 
   +C_F^2 ( \frac{331619149}{400075200} )
  + n_f C_F (  - \frac{12613}{238140} )   \Bigr] \\
& &  + a_s^3 \Bigl[
   C_F C_A^2 (  - \frac{840976971727}{129624364800}
                 - \frac{58649}{6615}\zeta_3 )
  + C_F^2 C_A  (  - \frac{16504689458671}{907370553600}
                     + \frac{58649}{2205} \zeta_3 ) \\
& & \hspace{5mm}
   + C_F^3 ( \frac{12876352060509647}{635159387520000}
               - \frac{117298}{6615} \zeta_3 )
   + n_f C_F C_A (  - \frac{3105820553}{6751269000} + \frac{296}{63}\zeta_3 ) \\
& & \hspace{5mm}
   + n_f C_F^2 ( \frac{8498139408671}{9073705536000} - \frac{296}{63}\zeta_3 )
   + n_f^2 C_F  ( \frac{339184373}{600112800} ) \Bigr] \\
& = & \hspace{3mm} a_s \Bigl(  -0.3915343915 \Bigr)  \\
& &+ a_s^2  \Bigl(  - 6.7576035061  - 0.0706195235 n_f \Bigr) \\
& &+ a_s^3  \Bigl( -134.7055041700 +12.3754453990 n_f
  +0.7536013741 n_f^2 \Bigr) \\
\end{array} $ 

\noindent $ \begin{array}{lll}
 \gamma_2^{G G}  &  = & \hspace{3mm} a_s n_f \frac{2}{3}  \\
& &  + a_s^2 \Bigl[
    n_f C_F ( \frac{74}{27} )
  + n_f C_A ( \frac{35}{27} )   \Bigr] \\
& &  + a_s^3 \Bigl[
     n_f C_F C_A (  - \frac{139}{9} + \frac{104}{3} \zeta_3 ) 
   + n_f C_F^2 ( \frac{2155}{243} - \frac{32}{3} \zeta_3 ) \\
& & \hspace{5mm}
   + n_f C_A^2  ( \frac{3589}{162} - 24 \zeta_3 ) 
   + n_f^2 C_F ( \frac{173}{243} )
   + n_f^2 C_A  (  - \frac{1058}{243} ) \Bigr] \\
& = & \hspace{3mm} a_s \Bigl( 0.6666666667  n_f \Bigr) \\
& &+ a_s^2  \Bigl( 7.5432098765 n_f \Bigr) \\
& &+ a_s^3  \Bigl( 37.6233727456 n_f-12.1124828532 n_f^2 \Bigr) \\
\end{array} $ \\
\vspace{.3cm}

\noindent $ \begin{array}{lll}
 \gamma_4^{G G}  &  = & \hspace{3mm} a_s 
     \Bigl[  C_A ( \frac{21}{5} ) + n_f  ( \frac{2}{3} ) \Bigr] \\
& &  + a_s^2 \Bigl[
  C_A^2 ( \frac{7121}{500} )   + n_f C_F ( \frac{20951}{9000} )
  + n_f C_A (  - \frac{2513}{450} )   \Bigr] \\
& &  + a_s^3 \Bigl[
   + C_A^3 ( \frac{103309639}{1350000} )
   + n_f C_F C_A (  - \frac{235535117}{4050000} + \frac{1632}{25}\zeta_3 ) \\
& & \hspace{5mm}
   + n_f C_F^2 ( \frac{2557151}{3037500} - \frac{66}{25} \zeta_3 )
   + n_f C_A^2  ( \frac{53797499}{2430000} - \frac{1566}{25} \zeta_3 )  \\
& & \hspace{5mm}
   + n_f^2 C_F (  - \frac{489887}{607500} )
   + n_f^2 C_A  (  - \frac{757861}{243000} )  \Bigr] \\
& = & \hspace{3mm} a_s  \Bigl(  12.6000000000 + 0.6666666667 n_f \Bigr)  \\
& &+ a_s^2  \Bigl(  128.1780000000  - 13.6494814815 n_f  \Bigr) \\
& &+ a_s^3  \Bigl( 2066.1927800000 -401.3127939259 n_f
  -10.4315064472 n_f^2 \Bigr) \\
\end{array} $ \\
\vspace{.3cm}

\noindent $ \begin{array}{lll}
 \gamma_6^{G G}  &  = & \hspace{3mm} a_s
     \Bigl[  C_A ( \frac{83}{14} )  + n_f ( \frac{2}{3} ) \Bigr] \\
& &  + a_s^2 \Bigl[
  C_A^2  ( \frac{1506899}{74088} )   + n_f C_F ( \frac{100319}{46305} )
  + n_f C_A (  - \frac{102997}{13230} )  \Bigr] \\
& &  + a_s^3 \Bigl[
   + C_A^3 ( \frac{96390174479}{871274880} )
   + n_f C_F C_A (  - \frac{9691228129}{130691232} 
                    + \frac{57256}{735}\zeta_3 ) \\
& & \hspace{5mm}
   + n_f C_F^2 (  - \frac{11024749151}{40841010000} - \frac{176}{147}\zeta_3 )
   + n_f C_A^2 ( \frac{9763460989}{466754400} - \frac{18792}{245} \zeta_3 ) \\
& & \hspace{5mm}
   + n_f^2 C_F (  - \frac{280414331}{291721500} )
   + n_f^2 C_A (  - \frac{6597737}{2083725} )   \Bigr] \\
& = & \hspace{3mm} a_s  \Bigl( 17.7857142857 + 0.6666666667 n_f \Bigr)  \\
& &+ a_s^2  \Bigl( 183.0538143829  - 20.4666846633 n_f   \Bigr) \\
& &+ a_s^3  \Bigl( 2987.0420583375  
         -566.6373298210 n_f -10.7806086102 n_f^2 \Bigr) \\
\end{array} $ 
\vspace{.3cm}

\noindent $ \begin{array}{lll}
 \gamma_8^{G G}  &  = & \hspace{3mm} a_s
     \Bigl[  C_A ( \frac{319}{45} )  + n_f ( \frac{2}{3} ) \Bigr] \\
& &  + a_s^2 \Bigl[
  C_A^2 ( \frac{2223694}{91125} )   + n_f C_F ( \frac{685883}{326592} )
  + n_f C_A (  - \frac{623687}{68040} )  \Bigr] \\
& &  + a_s^3 \Bigl[
   + C_A^3 ( \frac{1381390082227}{10333575000} )
   + n_f C_F C_A (  - \frac{220111823810087}{2592487296000} 
                    + \frac{81941}{945} \zeta_3 ) \\
& & \hspace{5mm}
   + n_f C_F^2 (  - \frac{14058417959723}{22684263840000}
                 - \frac{37}{54} \zeta_3 )
   + n_f C_A^2  ( \frac{2080130771161}{102876480000}
                  - \frac{162587}{1890} \zeta_3 )  \\
& & \hspace{5mm}
   + n_f^2 C_F (  - \frac{1747563703}{1687817250} )
   + n_f^2 C_A  (  - \frac{420970849}{128595600} )  \Bigr] \\ 
& = & \hspace{3mm} a_s  \Bigl( 21.2666666667 + 0.6666666667 n_f \Bigr)  \\
& &+ a_s^2  \Bigl( 219.6240987654 - 24.6992643216 n_f   \Bigr) \\
& &+ a_s^3  \Bigl( 3609.3541896322  
   -673.9430658122 n_f-11.2013383657 n_f^2 \Bigr) \\
\end{array} $ \\

\noindent $ \begin{array}{lll}
   C^{\psi}_{2,2} &  = &  1\\ & &
       + a_s C_F   ( \frac{1}{3} ) \\ & &
       + a_s^2 fl_{02} n_f C_F   (  - \frac{133}{81} ) \\ & &
       + a_s^2 n_f C_F   (  - 4 ) \\ & &
   + a_s^2 C_F C_A   ( \frac{3677}{135} - \frac{128}{5} \zeta_3 ) \\ & &
 + a_s^2 C_F^2   (  - \frac{4189}{810} + \frac{96}{5} \zeta_3 ) \\ & &
+ a_s^3 fl_{11} n_f C_F C_A 
  (  - \frac{186}{5} + \frac{152}{5} \zeta_3 + 32 \zeta_5 ) \\ & &
+ a_s^3 fl_{11} n_f C_F^2
    ( \frac{496}{5} - \frac{1216}{15} \zeta_3 - \frac{256}{3} \zeta_5 ) \\ & &
+ a_s^3 fl_{02} n_f C_F C_A
   ( \frac{1177679}{21870} - \frac{1376}{45} \zeta_3 - \frac{32}{3}
         \zeta_4 - \frac{64}{3} \zeta_5 ) \\ & &
+ a_s^3 fl_{02} n_f C_F^2 
  (  - \frac{28249}{2430} - \frac{17296}{405} \zeta_3 + \frac{32}{3}
         \zeta_4 + \frac{128}{3} \zeta_5 ) \\ & &
+ a_s^3 fl_{02} n_f^2 C_F
   (  - \frac{77623}{10935} + \frac{3232}{405} \zeta_3 ) \\ & &
+ a_s^3 n_f C_F C_A 
  (  - \frac{998153}{10935} + \frac{17432}{405} \zeta_3 - \frac{32}{3}
         \zeta_4 + \frac{80}{3} \zeta_5 ) \\ & &
+ a_s^3 n_f C_F^2 
  (  - \frac{341701}{21870} - \frac{1352}{45} \zeta_3 +
 \frac{32}{3} \zeta_4 ) \\ & &
+ a_s^3 n_f^2 C_F   ( \frac{7814}{2187} + \frac{64}{81} \zeta_3 ) \\ & &
+ a_s^3 C_F C_A^2  
 ( \frac{3667498}{10935} - \frac{46160}{81} \zeta_3 + \frac{32}{3} \zeta_4 +
         296 \zeta_5 ) \\ & &
+ a_s^3 C_F^2 C_A  
 (  - \frac{1013578}{10935} + \frac{30776}{45} \zeta_3 - 32 \zeta_4 -
         \frac{1552}{3} \zeta_5 ) \\ & &
+ a_s^3 C_F^3   (  - \frac{201577}{7290} - \frac{50864}{405} \zeta_3 
+ \frac{64}{3} \zeta_4 + \frac{416}{3} \zeta_5 )  \\ 
& = & 1 + a_s  \Bigl( 0.4444444444 \Bigr)   \\
& & + a_s^2  \Bigl(
 -2.1893004115 n_f fl_{02} +17.6937658911 -5.3333333333 n_f \Bigr)  \\
& & + a_s^3  \Bigl( 
  -24.0920133486 n_f fl_{11}
  -79.044861424 fl_{02} n_f
  + 3.3255044776 fl_{02} n_f^2  \\
& &  +442.7409692714 
  -165.1971095394 n_f 
  +6.0302724150 n_f^2 \Bigr)  
\end{array} $
\vspace{.3cm}

\noindent $ \begin{array}{lll}
 C^{\psi}_{2,4}  &  = &  1\\ & &
       + a_s C_F   ( \frac{91}{20} ) \\ & &
       + a_s^2 fl_{02} n_f C_F   ( \frac{393523}{1080000} ) \\ & &
   + a_s^2 n_f C_F   (  - \frac{1376021}{108000} ) \\ & &
   + a_s^2 C_F C_A   ( \frac{4649639}{54000} - \frac{269}{5} \zeta_3 ) \\ & &
 + a_s^2 C_F^2   (  - \frac{142624259}{6480000} + \frac{224}{5} \zeta_3 ) \\ & &
+ a_s^3 fl_{11} n_f C_F C_A 
  (  - \frac{154}{15} + \frac{76}{5} \zeta_3 + 16 \zeta_5 ) \\ & &
+ a_s^3 fl_{11} n_f C_F^2 
  ( \frac{1232}{45} - \frac{608}{15} \zeta_3 - \frac{128}{3} \zeta_5 ) \\ & &
+ a_s^3 fl_{02} n_f C_F C_A 
  ( \frac{20953881467}{1458000000} - \frac{10438}{1125} \zeta_3
          - \frac{121}{75} \zeta_4 ) \\ & &
 + a_s^3 fl_{02} n_f C_F^2  
 ( \frac{107812550749}{15309000000} - \frac{161386}{70875}
         \zeta_3 + \frac{121}{75} \zeta_4 ) \\ & &
+ a_s^3 fl_{02} n_f^2 C_F 
  (  - \frac{16205516}{6834375} + \frac{1204}{2025} \zeta_3 ) \\ & &
+ a_s^3 n_f C_F C_A 
  (  - \frac{36939762439}{122472000} + \frac{365878}{2835} \zeta_3
          - \frac{314}{15} \zeta_4 ) 
\end{array} $

\noindent $ \begin{array}{lll}
 \hspace{.8cm}  &  \hspace{.4cm} &
+ a_s^3 n_f C_F^2  
 (  - \frac{283057919183}{3061800000} - \frac{222433}{4725} \zeta_3
          + \frac{314}{15} \zeta_4 ) \\ & &
+ a_s^3 n_f^2 C_F 
  ( \frac{136926751}{8748000} + \frac{628}{405} \zeta_3 ) \\ & &
+ a_s^3 C_F C_A^2  
 ( \frac{26155194283}{24494400} - \frac{16471321}{14175} \zeta_3 +
         \frac{1439}{150} \zeta_4 + 470 \zeta_5 ) \\ & &
+ a_s^3 C_F^2 C_A   
(  - \frac{1209798320299}{12247200000} + \frac{21020311}{47250}
         \zeta_3 - \frac{1439}{50} \zeta_4 - \frac{124}{3} \zeta_5 ) \\ & &
+ a_s^3 C_F^3  
 ( \frac{4987373708399}{40824000000} + \frac{119936813}{283500} \zeta_3
          + \frac{1439}{75} \zeta_4 - \frac{1624}{3} \zeta_5 )  \\
& = & 1 + a_s  \Bigl(  6.0666666667 \Bigr)  \\
& & + a_s^2  \Bigl(
  0.4858308642  n_f fl_{02}
  +142.3434719201 
  -16.9879135802 n_f
 \Bigr)  \\
& & + a_s^3  \Bigl(
   -18.2188461805 n_f fl_{11}
   +16.6483484853 n_f fl_{02}
   -2.2086306890  n_f^2 fl_{02}
   \\ & & 
 +4169.2678883092 
 -901.2351625706 n_f
 +23.3550392440 n_f^2
   \Bigr) 
\end{array} $
\vspace{.3cm}

\noindent $ \begin{array}{lll} 
      C^{\psi}_{2,6} &  = & 1\\ & &
       + a_s C_F   ( \frac{5281}{630} ) \\ & &
       + a_s^2 fl_{02} n_f C_F   ( \frac{47344331}{129654000} ) \\ & &
       + a_s^2 n_f C_F   (  - \frac{1167429869}{55566000} ) \\ & &
      + a_s^2 C_F C_A  
 ( \frac{2962167319}{22226400} - \frac{348}{5} \zeta_3 ) \\ & &
      + a_s^2 C_F^2 
  (  - \frac{44524614749}{3889620000} + \frac{2036}{35} \zeta_3 ) \\ & &
      + a_s^3 fl_{11} n_f C_F C_A 
  (  - \frac{313979}{8400} + \frac{1028}{105} \zeta_3 + \frac{320}{7}
         \zeta_5 ) \\ & &
       + a_s^3 fl_{11} n_f C_F^2 
  ( \frac{313979}{3150} - \frac{8224}{315} \zeta_3 - \frac{2560}{21}
         \zeta_5 ) \\ & &
       + a_s^3 fl_{02} n_f C_F C_A   ( \frac{42613645803313}{5145967260000}
  - \frac{907166}{231525} \zeta_3 - \frac{484}{735} \zeta_4 ) \\ & &
       + a_s^3 fl_{02} n_f C_F^2   ( \frac{44744425895431}{5145967260000}
 - \frac{1313258}{694575} \zeta_3 + \frac{484}{735} \zeta_4 ) \\ & &
       + a_s^3 fl_{02} n_f^2 C_F
   (  - \frac{18418888183}{13613670000} + \frac{3448}{19845}
         \zeta_3 )  \\  & &
       + a_s^3 n_f C_F C_A  
 (  - \frac{109033584779183}{210039480000} + \frac{358312}{2025}
         \zeta_3 - \frac{2836}{105} \zeta_4 ) \\ & &
       + a_s^3 n_f C_F^2 
  (  - \frac{11225895261923}{51051262500} - \frac{417224}{6615}
         \zeta_3 + \frac{2836}{105} \zeta_4 ) \\ & &
       + a_s^3 n_f^2 C_F 
  ( \frac{11510197169}{388962000} + \frac{5672}{2835} \zeta_3 ) \\ & &
       + a_s^3 C_F C_A^2 
  ( \frac{492030789922433}{280052640000} - \frac{163363328}{99225}
         \zeta_3 + \frac{34931}{3675} \zeta_4 + \frac{4436}{7} \zeta_5 ) \\ & &
       + a_s^3 C_F^2 C_A 
  ( \frac{86882592642607}{204205050000} + \frac{174615691}{385875}
         \zeta_3 - \frac{34931}{1225} \zeta_4 - \frac{1116}{7} \zeta_5 ) \\ & &
       + a_s^3 C_F^3 
  ( \frac{4962846338797159}{257298363000000} + \frac{2402912488}{3472875}
          \zeta_3 + \frac{69862}{3675} \zeta_4 - \frac{4168}{7} \zeta_5 )  \\ 
& = & 1 + a_s  \Bigl(  11.1767195767   \Bigr)   \\
& & + a_s^2  \Bigl(
  0.4868787285 n_f fl_{02}
  +302.3987349886 
  -28.0130504025 n_f
 \Bigr)  \\
&  & + a_s^3  \Bigl(
 -16.1427176094 n_f fl_{11}
 +24.1177881298 n_f fl_{02}
 -1.5254891426 n_f^2 fl_{02}
 \\ & &
  +10069.6308450937 
  -1816.3229292502 n_f
  +42.6627311588 n_f^2
 \Bigr) 
\end{array} $
\vspace{.3cm}

\noindent $ \begin{array}{lll}
       C^{\psi}_{2,8} &  = & 1\\ & &
       + a_s C_F   ( \frac{58703}{5040} ) \\ & &
       + a_s^2 fl_{02} n_f C_F   ( \frac{83382717493}{288054144000} ) \\ & &
       + a_s^2 n_f C_F   (  - \frac{682775147983}{24004512000} ) \\ & &
    + a_s^2 C_F C_A 
  ( \frac{3348220230689}{19203609600} - \frac{17071}{210} \zeta_3 ) \\ & &
   + a_s^2 C_F^2 
  ( \frac{203744735078509}{20163790080000} + \frac{2396}{35} \zeta_3 ) \\ & &
       + a_s^3 fl_{11} n_f C_F C_A   (  - \frac{476784977}{8930250}
  + \frac{36443}{9450} \zeta_3
          + \frac{200}{3} \zeta_5 ) \\ & &
   + a_s^3 fl_{11} n_f C_F^2 
  ( \frac{1907139908}{13395375} - \frac{145772}{14175} \zeta_3 -
         \frac{1600}{9} \zeta_5 ) \\ & &
    + a_s^3 fl_{02} n_f C_F C_A   ( \frac{91829623403140453}{16332669964800000} 
    - \frac{23440003}{10716300} \zeta_3 - \frac{1369}{3780} \zeta_4 ) \\ & &
       + a_s^3 fl_{02} n_f C_F^2  
  ( \frac{10181295454316382631}{1257615587289600000}
  - \frac{174416989}{117879300} \zeta_3 + \frac{1369}{3780} \zeta_4 ) \\ & &
       + a_s^3 fl_{02} n_f^2 C_F   (  - \frac{16959835520839}{19443654720000} 
 + \frac{2017}{25515} \zeta_3 ) \\ & &
       + a_s^3 n_f C_F C_A   (  - \frac{4344582726190968061}{5988645653760000} +
         \frac{562955509}{2619540} \zeta_3 - \frac{9883}{315} \zeta_4 ) \\ & &
       + a_s^3 n_f C_F^2 
  (  - \frac{119458136235341463641}{314403896822400000} -
         \frac{41139607}{523908} \zeta_3 + \frac{9883}{315} \zeta_4 ) \\ & &
       + a_s^3 n_f^2 C_F 
  ( \frac{1187935000932371}{27221116608000} + \frac{19766}{8505}
         \zeta_3 ) \\ & &
       + a_s^3 C_F C_A^2   ( \frac{327176399149038763}{136105583040000}
  - \frac{1614116849}{793800} \zeta_3  
         + \frac{2510407}{264600} \zeta_4 + \frac{47044}{63} \zeta_5 ) \\ & &
     + a_s^3 C_F^2 C_A   ( \frac{5717961692522254249661}{5030462349158400000} +
         \frac{339247197811}{916839000} \zeta_3 
            - \frac{2510407}{88200} \zeta_4 - \frac{1982}{9} \zeta_5) \\ & &
    + a_s^3 C_F^3   (  - \frac{14319995334117562552891}{352132364441088000000} 
     + \frac{33131600990227}{33006204000} \zeta_3 
 + \frac{2510407}{132300} \zeta_4 - 664
         \zeta_5 )  \\ 
& = & 1 + a_s  \Bigl( 15.5298941799  \Bigr)   \\
& & + a_s^2  \Bigl( 
 0.3859585393 n_f fl_{02}
 +470.8074190065 
 -37.9248227990 n_f
 \Bigr)  \\
& & + a_s^3  \Bigl(
   -15.0920382729 n_f fl_{11}
   +22.3320193843 n_f fl_{02}
   -1.0363081217 n_f^2 fl_{02}
 \\ & &
 +17162.3724471532 
 -2787.2976921073 n_f
 +61.9117799688 n_f^2
 \Bigr) 
\end{array} $
\vspace{.3cm}

\noindent $ \begin{array}{lll}
 C^{G}_{2,2} &  = &
      \hspace{3mm} a_s n_f   (  - \frac{1}{2} ) \\ & &
       + a_s^2 n_f C_F   (  - \frac{4799}{810} + \frac{16}{5} \zeta_3 ) \\ & &
       + a_s^2 n_f C_A   ( \frac{115}{324} - 2 \zeta_3 ) \\ & &
 + a_s^3 fl^g_{11} n_f^2 C_F 
  (  - 4 + \frac{272}{15} \zeta_3 - \frac{64}{3} \zeta_5 ) \\ & &
    + a_s^3 fl^g_{11} n_f^2 C_A 
  ( \frac{3}{2} - \frac{34}{5} \zeta_3 + 8 \zeta_5 ) \\ & &
      + a_s^3 n_f C_F C_A  
 (  - \frac{161284}{3645} + \frac{274}{15} \zeta_3 - \frac{52}{3} \zeta_4 +
         40 \zeta_5 ) \\ & &
       + a_s^3 n_f C_F^2  
 ( \frac{28403}{4860} + \frac{4148}{405} \zeta_3 + \frac{16}{3} \zeta_4 
 \\ & & + a_s^3 n_f C_A^2  
 (  - \frac{3444493}{87480} + \frac{1828}{405} \zeta_3 + 12 \zeta_4 +
         12 \zeta_5 ) \\ & &
       + a_s^3 n_f^2 C_F 
  ( \frac{23291}{4860} - \frac{104}{405} \zeta_3 ) \\ & &
       + a_s^3 n_f^2 C_A 
  ( \frac{130219}{21870} + \frac{622}{405} \zeta_3 )  \\
\end{array} $

\noindent $ \begin{array}{lll}
\hspace{.8cm} & = & a_s  \Bigl(  - 0.5 n_f \Bigr)   \\
& & + a_s^2  \Bigl(
   - 8.9183389614 n_f
 \Bigr)  \\
& & + a_s^3  \Bigl(
 -0.9007972776 n_f^2 fl^g_{11}
 -130.7340963277 n_f
 +29.3793351486 n_f^2
 \Bigr) 
\end{array} $

\noindent $ \begin{array}{lll}
     C^{G}_{2,4}  & = & \hspace{3mm}
        a_s n_f   (  - \frac{133}{180} ) \\ & &
       + a_s^2 n_f C_F  
 (  - \frac{6410867}{720000} + \frac{18}{5} \zeta_3 ) \\ & &
       + a_s^2 n_f C_A  
 (  - \frac{1812647}{3240000} - \frac{9}{5} \zeta_3 ) \\ & &
+ a_s^3 fl^g_{11} n_f^2 C_F( - \frac{410}{9} - \frac{10592}{45} \zeta_3 
 + \frac{928}{3} \zeta_5 ) \\ & &
+ a_s^3 fl^g_{11} n_f^2 C_A 
  ( \frac{205}{12} + \frac{1324}{15} \zeta_3 - 116 \zeta_5 ) \\ & &
      + a_s^3 n_f C_F C_A  
 (  - \frac{1571010034879}{61236000000} - \frac{25747}{630} \zeta_3
          - \frac{969}{50} \zeta_4 + 82 \zeta_5 ) \\ & &
       + a_s^3 n_f C_F^2 
  (  - \frac{4566936087251}{122472000000} + \frac{18867971}{283500}
         \zeta_3 + \frac{261}{50} \zeta_4 - \frac{292}{3} \zeta_5 ) \\ & &
       + a_s^3 n_f C_A^2 
  (  - \frac{77953723693}{729000000} + \frac{1312997}{20250} \zeta_3
          + \frac{354}{25} \zeta_4 - \frac{28}{3} \zeta_5 ) \\ & &
+a_s^3 n_f^2 C_F( \frac{790842890021}{61236000000}
  - \frac{595883}{141750} \zeta_3 ) \\ & &
       + a_s^3 n_f^2 C_A   ( \frac{109146757}{9112500}
  + \frac{59}{2025} \zeta_3 )  \\
& = & a_s  \Bigl(    - 0.7388888889 n_f \Bigr)   \\
& & + a_s^2  \Bigl(  - 14.2715869197 n_f
 \Bigr)  \\
& & + a_s^3  \Bigl(
  -1.6118165124 n_f^2 fl^g_{11}
 -346.4612756184 n_f
 +46.5201756441 n_f^2
 \Bigr) 
\end{array} $
\vspace{.3cm}

\noindent $ \begin{array}{lll}
     C^{G}_{2,6} &  = &   \hspace{3mm}
       a_s n_f   (  - \frac{1777}{2520} ) \\ & &
       + a_s^2 n_f C_F  
 (  - \frac{12660217}{1296540} + \frac{20}{7} \zeta_3 ) \\ & &
       + a_s^2 n_f C_A  
 (  - \frac{16794343}{7779240} - \frac{10}{7} \zeta_3 ) \\ & &
 + a_s^3 fl^g_{11} n_f^2 C_F  
 (  - \frac{1330183}{9072} - \frac{1673}{3} \zeta_3 + \frac{16400}{21}
         \zeta_5 ) \\ & &
       + a_s^3 fl^g_{11} n_f^2 C_A  
 ( \frac{1330183}{24192} + \frac{1673}{8} \zeta_3 - \frac{2050}{7}
         \zeta_5 ) \\ & &
       + a_s^3 n_f C_F C_A   (  - \frac{395582364061093}{13722579360000}
  - \frac{4535294}{231525} \zeta_3 - \frac{11833}{735} \zeta_4 
 + \frac{1040}{21} \zeta_5 ) \\ & &
       + a_s^3 n_f C_F^2   (  - \frac{986797608696253}{13722579360000}
  + \frac{65099719}{1389150} \zeta_3 + \frac{620}{147} \zeta_4
  - \frac{440}{7} \zeta_5 ) \\ & &
       + a_s^3 n_f C_A^2   (  - \frac{230534380633807}{1646709523200}
  + \frac{43271551}{694575} \zeta_3 + \frac{2911}{245} \zeta_4
  - \frac{80}{21} \zeta_5 ) \\ & &
       + a_s^3 n_f^2 C_F  
 ( \frac{8883735865961}{571774140000} - \frac{485917}{138915} \zeta_3
          ) \\ & &
 + a_s^3 n_f^2 C_A 
  ( \frac{4519688643989}{294055272000} + \frac{143}{7938} \zeta_3 )  \\
& = & a_s  \Bigl(  - 0.7051587302 n_f \Bigr)   \\
& & + a_s^2  \Bigl(   - 20.0684982842 n_f
 \Bigr)  \\
& & + a_s^3  \Bigl(
   -1.4960369382 n_f^2 fl^g_{11}
   -715.0372438398 n_f
   +61.2854509604 n_f^2
 \Bigr) 
\end{array} $
\vspace{.3cm}

\noindent $ \begin{array}{lll}
   C^{G}_{2,8} &  =  & \hspace{3mm}
        a_s n_f   (  - \frac{16231}{25200} ) \\ & &
       + a_s^2 n_f C_F 
  (  - \frac{23994462871}{2351462400} + \frac{7}{3} \zeta_3 ) \\ & &
       + a_s^2 n_f C_A  
 (  - \frac{89216988167}{29393280000} - \frac{7}{6} \zeta_3 ) \\ & &
 + a_s^3 fl^g_{11} n_f^2 C_F 
  (  - \frac{692580563}{1913625} - \frac{2357104}{2025} \zeta_3
          + \frac{15232}{9} \zeta_5 ) \\ & &
    + a_s^3 fl^g_{11} n_f^2 C_A 
  ( \frac{692580563}{5103000} + \frac{294638}{675} \zeta_3 -
         \frac{1904}{3} \zeta_5 ) \\ & &
  + a_s^3 n_f C_F C_A   (  - \frac{13175786598173293133}{359318739225600000} -
 \frac{2254405091}{336798000} \zeta_3 
  - \frac{515201}{37800} \zeta_4 + \frac{289}{9} \zeta_5 ) \\ & &
       + a_s^3 n_f C_F^2 
  (  - \frac{954577609184015777449}{10060924698316800000} +
  \frac{154197212633}{4715172000} \zeta_3 
 + \frac{749}{216} \zeta_4 - \frac{130}{3} \zeta_5 ) \\ & &
       + a_s^3 n_f C_A^2   (  - \frac{3720425499755558929}{23332385664000000} +
  \frac{12414318401}{214326000} \zeta_3 
   + \frac{64021}{6300} \zeta_4 - \frac{10}{9} \zeta_5 ) \\ & &
       + a_s^3 n_f^2 C_F   ( \frac{4082506317130294583}{239545826150400000} -
         \frac{187698097}{67359600} \zeta_3 ) \\ & &
    + a_s^3 n_f^2 C_A  
 ( \frac{3281715651209153}{194436547200000} - \frac{4591}{72900}
         \zeta_3 )  \\
& = & a_s  \Bigl(   - 0.6440873016  n_f \Bigr)   \\
& & + a_s^2  \Bigl(   - 23.1787352382 n_f
 \Bigr)  \\
& & + a_s^3  \Bigl(
  -1.2864009150 n_f^2 fl^g_{11}
  -996.5038709496 n_f
  +68.6646730444 n_f^2
 \Bigr) 
\end{array} $
\vspace{.3cm}

\noindent $ \begin{array}{lll}
    C^{\psi}_{L,2} &  = & \hspace{3mm}
        a_s C_F   ( \frac{4}{3} ) \\ & &
       + a_s^2 fl_{02} n_f C_F   (  - \frac{80}{27} ) \\ & &
       + a_s^2 n_f C_F   (  - \frac{92}{27} ) \\ & &
       + a_s^2 C_F C_A   ( \frac{2878}{135} - \frac{32}{5} \zeta_3 ) \\ & &
       + a_s^2 C_F^2   (  - \frac{1906}{135} + \frac{64}{5} \zeta_3 ) \\ & &
 + a_s^3 fl_{11} n_f C_F C_A (  - \frac{164}{5} 
+ \frac{1008}{5} \zeta_3 - 192 \zeta_5 ) \\ & &
+ a_s^3 fl_{11} n_f C_F^2 ( \frac{1312}{15} - \frac{2688}{5} \zeta_3 
+ 512 \zeta_5 ) \\ & &
       + a_s^3 fl_{02} n_f C_F C_A  
 (  - \frac{48058}{1215} + \frac{304}{5} \zeta_3 - \frac{256}{3}
         \zeta_5 ) \\ & &
       + a_s^3 fl_{02} n_f C_F^2   
(  - \frac{46898}{1215} - \frac{5056}{45} \zeta_3 + \frac{512}{3}
         \zeta_5 ) \\ & &
      + a_s^3 fl_{02} n_f^2 C_F 
  ( \frac{3364}{405} + \frac{64}{15} \zeta_3 ) \\ & &
     + a_s^3 n_f C_F C_A   (  - \frac{204548}{1215} - \frac{1568}{45} \zeta_3 
+ \frac{320}{3} \zeta_5
       ) \\ & &
    + a_s^3 n_f C_F^2   ( \frac{25534}{405} - \frac{2848}{45} \zeta_3 ) \\ & &
   + a_s^3 n_f^2 C_F   ( \frac{2168}{243} ) \\ & &
  + a_s^3 C_F C_A^2  
 ( \frac{548668}{1215} - \frac{3680}{9} \zeta_3 + 224 \zeta_5 ) \\ & &
 + a_s^3 C_F^2 C_A 
(  - \frac{41536}{405} + \frac{73504}{45} \zeta_3 - 
 \frac{5248}{3} \zeta_5 ) \\ & &
 + a_s^3 C_F^3 (  - \frac{232798}{1215} - \frac{39424}{45} \zeta_3 
 + \frac{3584}{3} \zeta_5 )  \\
& = & a_s  \Bigl( 1.7777777778  \Bigr)   \\
& & + a_s^2  \Bigl( 
 -3.9506172840 n_f fl_{02}
 +56.7553015166 
 -4.5432098765 n_f 
 \Bigr)  \\
& & + a_s^3  \Bigl(
 -7.7366982885 n_f fl_{11}
 -213.9253075658  n_f fl_{02}
 +17.9132652795 n_f^2 fl_{02}
  \\ & &
 +2544.5980873744 
 -421.6908884762 n_f
 +11.8957475995 n_f^2
 \Bigr) 
\end{array} $
\vspace{.3cm}

\noindent $ \begin{array}{lll}
   C^{\psi}_{L,4} &  = & \hspace{3mm}
        a_s C_F   ( \frac{4}{5} ) \\ & &
       + a_s^2 fl_{02} n_f C_F   (  - \frac{586}{1125} ) \\ & &
       + a_s^2 n_f C_F   (  - \frac{64}{25} ) \\ & &
       + a_s^2 C_F C_A   ( \frac{4763}{225} - \frac{48}{5} \zeta_3 ) \\ & &
       + a_s^2 C_F^2   (  - \frac{19967}{1125} + \frac{96}{5} \zeta_3 ) \\ & &
 + a_s^3 fl_{11} n_f C_F C_A 
( \frac{12359}{90} + \frac{856}{15} \zeta_3 - 192 \zeta_5 ) \\ & &
 + a_s^3 fl_{11} n_f C_F^2 
(  - \frac{49436}{135} - \frac{6848}{45} \zeta_3 + 512 \zeta_5 )
   \\ & &
 + a_s^3 fl_{02} n_f C_F C_A 
(  - \frac{5074537}{405000} - \frac{824}{675} \zeta_3 ) \\ & &
 + a_s^3 fl_{02} n_f C_F^2 
(  - \frac{7715012}{1063125} + \frac{29488}{4725} \zeta_3 ) \\ & &
 + a_s^3 fl_{02} n_f^2 C_F   ( \frac{106981}{60750} ) \\ & &
 + a_s^3 n_f C_F C_A  
 (  - \frac{14259893}{94500} + \frac{55904}{945} \zeta_3 ) \\ & &
 + a_s^3 n_f C_F^2 
  ( \frac{258828431}{2835000} - \frac{33344}{315} \zeta_3 ) \\ & &
 + a_s^3 n_f^2 C_F   ( \frac{82688}{10125} ) \\ & &
 + a_s^3 C_F C_A^2 
  ( \frac{171354151}{283500} - \frac{1889656}{4725} \zeta_3 + 32
         \zeta_5 ) \\ & &
 + a_s^3 C_F^2 C_A 
  (  - \frac{263016326}{354375} + \frac{1070306}{1575} \zeta_3 + 32
         \zeta_5 ) \\ & &
 + a_s^3 C_F^3 ( \frac{582157141}{3780000} 
 + \frac{57356}{945} \zeta_3 - 192 \zeta_5 )  \\
& = & a_s  \Bigl( 1.0666666667  \Bigr)  \\
& & + a_s^2  \Bigl( 
  -0.6945185185 n_f fl_{02}
  +47.9939893120 
  -3.4133333333 n_f
 \Bigr)  \\
& & + a_s^3  \Bigl(
  -5.0588695123 n_f fl_{11}
  -55.5530455971 n_f fl_{02}
  +2.3480054870 n_f^2 fl_{02}
  \\ & &
  +2523.7390200791
  -383.0520013416 n_f
  +10.8889547325 n_f^2
 \Bigr) 
\end{array} $
\vspace{.3cm}

\noindent $ \begin{array}{lll} 
   C^{\psi}_{L,6}  &  = & \hspace{3mm}
        a_s C_F   ( \frac{4}{7} ) \\ & &
       + a_s^2 fl_{02} n_f C_F   (  - \frac{14614}{77175} ) \\ & &
       + a_s^2 n_f C_F   (  - \frac{1486}{735} ) \\ & &
       + a_s^2 C_F C_A   ( \frac{172106}{11025} - \frac{48}{7} \zeta_3 ) \\ & &
       + a_s^2 C_F^2   (  - \frac{257318}{25725} + \frac{96}{7} \zeta_3 ) \\ & &
       + a_s^3 fl_{11} n_f C_F C_A 
  ( \frac{5869993}{63000} + \frac{944}{21} \zeta_3 - \frac{960}{7}
         \zeta_5 ) \\ & &
     + a_s^3 fl_{11} n_f C_F^2  
 (  - \frac{5869993}{23625} - \frac{7552}{63} \zeta_3 + \frac{2560}{7}
         \zeta_5 ) \\ & &
       + a_s^3 fl_{02} n_f C_F C_A 
  (  - \frac{41368883483}{8168202000} - \frac{7384}{11025}
         \zeta_3 ) \\ & &
  + a_s^3 fl_{02} n_f C_F^2  
 (  - \frac{193467963407}{61261515000} + \frac{416}{175} \zeta_3
          ) \\ & &
       + a_s^3 fl_{02} n_f^2 C_F   ( \frac{27905687}{48620250} ) \\ & &
  + a_s^3 n_f C_F C_A  
 (  - \frac{927531079}{6945750} + \frac{88916}{1575} \zeta_3 ) \\ & &
       + a_s^3 n_f C_F^2  
 ( \frac{66489992539}{875164500} - \frac{31312}{315} \zeta_3 ) \\ & &
       + a_s^3 n_f^2 C_F   ( \frac{986872}{138915} )
\end{array} $

\noindent $ \begin{array}{lll}
\hspace{.8cm} & & 
       + a_s^3 C_F C_A^2   ( \frac{3186722339}{6174000}
   - \frac{4977821}{11025} \zeta_3 + \frac{1040}{7} \zeta_5 ) \\ & &
       + a_s^3 C_F^2 C_A   
(  - \frac{4333877446411}{8751645000} + \frac{2071508}{2205} \zeta_3
          - 480 \zeta_5 ) \\ & &
       + a_s^3 C_F^3   (  - \frac{90265366481}{2268945000}
  - \frac{536548}{2205} \zeta_3 + \frac{2560}{7} \zeta_5 ) \\
 & = & a_s  \Bigl( 0.7619047619 \Bigr)   \\
& & + a_s^2  \Bigl( 
 -0.2524824533 n_f fl_{02}
 +40.9961975991 
 -2.6956916100 n_f
 \Bigr)  \\
& & + a_s^3  \Bigl(
  -3.7056125257 n_f fl_{11}
  -24.0132253893 n_f fl_{02}
  +0.7652692585 n_f^2 fl_{02}
  \\ & &
  +2368.1937754336 
  -340.0691069253 n_f
  +9.4721904282 n_f^2
 \Bigr) 
\end{array} $
\vspace{.3cm}

\noindent $ \begin{array}{lll}
   C^{\psi}_{L,8} &  = & \hspace{3mm}
       a_s C_F   ( \frac{4}{9} ) \\ & &
       + a_s^2 fl_{02} n_f C_F   (  - \frac{6523}{71442} ) \\ & &
       + a_s^2 n_f C_F   (  - \frac{14234}{8505} ) \\ & &
       + a_s^2 C_F C_A  
 ( \frac{14741729}{1190700} - \frac{16}{3} \zeta_3 ) \\ & &
       + a_s^2 C_F^2  
 (  - \frac{21694349}{3572100} + \frac{32}{3} \zeta_3 ) \\ & &
  + a_s^3 fl_{11} n_f C_F C_A 
  ( \frac{35555777437}{500094000} + \frac{170756}{4725} \zeta_3
          - \frac{320}{3} \zeta_5 ) \\ & &
       + a_s^3 fl_{11} n_f C_F^2  
 (  - \frac{35555777437}{187535250} - \frac{1366048}{14175}
         \zeta_3 + \frac{2560}{9} \zeta_5 ) \\ & &
       + a_s^3 fl_{02} n_f C_F C_A   (  - \frac{4264597461073}{1620304560000}
  - \frac{12892}{33075} \zeta_3 ) \\ & &
       + a_s^3 fl_{02} n_f C_F^2   (  - \frac{55595717822203}{31190862780000}
  + \frac{1327432}{1091475} \zeta_3 ) \\ & &
       + a_s^3 fl_{02} n_f^2 C_F   ( \frac{3487324703}{13502538000} ) \\ & &
       + a_s^3 n_f C_F C_A   (  - \frac{23153083641529}{198037224000} 
  + \frac{18459136}{363825} \zeta_3 ) \\ & &
       + a_s^3 n_f C_F^2 
  ( \frac{11844644404289}{198037224000} - \frac{914992}{10395} \zeta_3
          ) \\ & &
       + a_s^3 n_f^2 C_F   ( \frac{1435876}{229635} ) \\ & &
       + a_s^3 C_F C_A^2  
 ( \frac{7653142193467}{18003384000} - \frac{9508318}{19845} \zeta_3
          + \frac{2240}{9} \zeta_5 ) \\ & &
       + a_s^3 C_F^2 C_A   (  - \frac{7667007621800089}{27725211360000}
  + \frac{2476882549}{2182950} \zeta_3 - \frac{2720}{3} \zeta_5 ) \\ & &
       + a_s^3 C_F^3   (  - \frac{86167166469418457}{499053804480000}
  - \frac{111668693}{218295} \zeta_3 + \frac{7360}{9} \zeta_5 )  \\ 
& = & a_s  \Bigl(  0.5925925926 \Bigr)   \\
& & + a_s^2  \Bigl( 
   -0.1217397796 n_f fl_{02}
   +35.8766440564 
   -2.2314716833  n_f
 \Bigr)  \\
& & + a_s^3  \Bigl(
    -2.9137025628 n_f fl_{11}
   -12.9718526710 n_f fl_{02}
   +0.3443623910 n_f^2 fl_{02}
  \\ & &
  +2215.2108750618 
  -305.4730328944 n_f
  +8.3371495344 n_f^2
 \Bigr) 
\end{array} $
\vspace{.3cm}

\noindent $ \begin{array}{lll}
   C^{G}_{L,2}  & = & \hspace{3mm}
       a_s n_f   ( \frac{2}{3} ) \\ & &
       + a_s^2 n_f C_F   (  - \frac{116}{135} - \frac{16}{5} \zeta_3 ) \\ & &
       + a_s^2 n_f C_A   ( \frac{173}{27} ) \\ & &
 + a_s^3 fl^g_{11} n_f^2 C_F  
 ( \frac{8}{3} + \frac{3808}{15} \zeta_3 - \frac{896}{3} \zeta_5 ) \\ & &
 + a_s^3 fl^g_{11} n_f^2 C_A  
 (  - 1 - \frac{476}{5} \zeta_3 + 112 \zeta_5 ) \\ & &
 + a_s^3 n_f C_F C_A   (  - \frac{71657}{1215} - \frac{248}{5} \zeta_3 
+ \frac{80}{3} \zeta_5 ) \\ & &
 + a_s^3 n_f C_F^2   ( \frac{51283}{1215} 
+ \frac{928}{45} \zeta_3 - \frac{160}{3} \zeta_5 ) \\ & &
   + a_s^3 n_f C_A^2   ( \frac{235283}{2430} - \frac{148}{5} \zeta_3 
+ \frac{64}{3} \zeta_5 ) \\ & &
   + a_s^3 n_f^2 C_F   ( \frac{9031}{1215} + \frac{256}{45} \zeta_3 ) \\ & &
   + a_s^3 n_f^2 C_A   (  - \frac{5431}{405} + \frac{4}{15} \zeta_3 )  \\
& = & a_s  \Bigl( 0.6666666667 n_f \Bigr)   \\
& & + a_s^2  \Bigl( 12.9477670897 n_f
 \Bigr)  \\
& & + a_s^3  \Bigl(
 -0.3889396640  n_f^2 fl^g_{11}
 +407.2806319596  n_f
 -20.2395974792 n_f^2
 \Bigr) 
\end{array} $
\vspace{.3cm}

\noindent $ \begin{array}{lll}
   C^{G}_{L,4}  & = & \hspace{3mm}
        a_s n_f   ( \frac{4}{15} ) \\ & &
       + a_s^2 n_f C_F   (  - \frac{2764}{1125} ) \\ & &
       + a_s^2 n_f C_A   ( \frac{19229}{3375} ) \\ & &
       + a_s^3 fl^g_{11} n_f^2 C_F  
 (  - \frac{5528}{135} + \frac{4384}{135} \zeta_3 ) \\ & &
       + a_s^3 fl^g_{11} n_f^2 C_A  
 ( \frac{691}{45} - \frac{548}{45} \zeta_3 ) \\ & &
       + a_s^3 n_f C_F C_A 
  (  - \frac{261555163}{1890000} - \frac{272858}{4725} \zeta_3 + 128
          \zeta_5 ) \\ & &
       + a_s^3 n_f C_F^2  
 ( \frac{912595079}{11340000} + \frac{225908}{4725} \zeta_3 - 128
         \zeta_5 ) \\ & &
   + a_s^3 n_f C_A^2  
 ( \frac{172349813}{1215000} + \frac{5036}{675} \zeta_3 - 32 \zeta_5 )
   \\ & &
       + a_s^3 n_f^2 C_F 
  ( \frac{4921307}{680400} - \frac{32}{105} \zeta_3 ) \\ & &
       + a_s^3 n_f^2 C_A 
  (  - \frac{445666}{30375} - \frac{8}{15} \zeta_3 )  \\
& = &  a_s  \Bigl(  0.2666666667  n_f \Bigr)   \\
& & + a_s^2  \Bigl( 13.8165925926 n_f
 \Bigr)  \\
& & + a_s^3  \Bigl(
  -0.3984298404 n_f^2 fl^g_{11}
  +767.7125420910  n_f
  -36.7841923177 n_f^2
 \Bigr) 
\end{array} $
\vspace{.3cm}

\noindent $ \begin{array}{lll}
     C^{G}_{L,6}  & = & \hspace{3mm}
        a_s n_f   ( \frac{1}{7} ) \\ & &
       + a_s^2 n_f C_F   (  - \frac{137761}{92610} ) \\ & &
       + a_s^2 n_f C_A   ( \frac{503977}{123480} ) \\ & &
 + a_s^3 fl^g_{11} n_f^2 C_F  
 ( \frac{7994213}{340200} + \frac{61778}{315} \zeta_3 - \frac{1760}{7}
         \zeta_5 ) \\ & &
 + a_s^3 fl^g_{11} n_f^2 C_A   (  - \frac{7994213}{907200}
 - \frac{30889}{420} \zeta_3 + \frac{660}{7} \zeta_5 ) \\ & &
       + a_s^3 n_f C_F C_A 
  (  - \frac{1620883880209}{16336404000} - \frac{51181}{2205}
         \zeta_3 + \frac{480}{7} \zeta_5 ) \\ & &
       + a_s^3 n_f C_F^2   ( \frac{101548438631}{2042050500}
  + \frac{43574}{2205} \zeta_3 - \frac{480}{7} \zeta_5 ) 
\end{array} $

\noindent $ \begin{array}{lll}
\hspace{.8cm} & &
       + a_s^3 n_f C_A^2   ( \frac{380381802767}{3267280800} 
 + \frac{10817}{4410} \zeta_3 - \frac{120}{7} \zeta_5 ) \\ & &
       + a_s^3 n_f^2 C_F  
 ( \frac{60190474091}{12252303000} - \frac{32}{63} \zeta_3 ) \\ & &
       + a_s^3 n_f^2 C_A 
  (  - \frac{847812559}{77792400} - \frac{2}{7} \zeta_3 )  \\
& = & a_s  \Bigl(  0.1428571429 n_f \Bigr)   \\
& & + a_s^2  \Bigl(   10.2609536407 n_f
 \Bigr)  \\
& & + a_s^3  \Bigl(
   -0.3055276256 n_f^2 fl^g_{11}
   +694.5092121136 n_f
   -27.9895081033 n_f^2
 \Bigr) 
\end{array} $
\vspace{.3cm}

\noindent $ \begin{array}{lll}
   C^{G}_{L,8}  & = & \hspace{3mm}
        a_s n_f   ( \frac{4}{45} ) \\ & &
       + a_s^2 n_f C_F   (  - \frac{51097}{51030} ) \\ & &
       + a_s^2 n_f C_A   ( \frac{7712869}{2551500} ) \\ & &
 + a_s^3 fl^g_{11} n_f^2 C_F 
  ( \frac{3665714041}{35721000} + \frac{617672}{1575} \zeta_3 -
         \frac{1664}{3} \zeta_5 ) \\ & &
     + a_s^3 fl^g_{11} n_f^2 C_A  
 (  - \frac{3665714041}{95256000} - \frac{77209}{525} \zeta_3
          + 208 \zeta_5 ) \\ & &
      + a_s^3 n_f C_F C_A   (  - \frac{520855237960033}{7129340064000}
 - \frac{6119609}{519750} \zeta_3 + \frac{128}{3} \zeta_5 ) \\ & &
       + a_s^3 n_f C_F^2 
  ( \frac{2384408424295187}{71293400640000} + \frac{7723411}{779625}
         \zeta_3 - \frac{128}{3} \zeta_5 ) \\ & &
       + a_s^3 n_f C_A^2  
 ( \frac{27404278602137}{289340100000} + \frac{20438}{23625} \zeta_3
          - \frac{32}{3} \zeta_5 ) \\ & &
       + a_s^3 n_f^2 C_F  
 ( \frac{124374980290567}{35646700320000} - \frac{608}{1485} \zeta_3
          ) \\ & &
       + a_s^3 n_f^2 C_A   
(  - \frac{11324757281}{1377810000} - \frac{8}{45} \zeta_3 ) \\
& = & a_s  \Bigl(  0.0888888889 n_f  \Bigr)   \\
& & + a_s^2  \Bigl( 7.7335451042 n_f 
 \Bigr)  \\
& & + a_s^3  \Bigl(
   -0.2322211886 n_f^2 fl^g_{11}
   +592.3307972098 n_f
   -21.3033368117 n_f^2
 \Bigr) 
\end{array} $
\[ \]
\noindent In addition to the results given above for the moments N=2,4,6,8,
 we present the following results for the non-singlet contributions 
 to the 10th moment.
\[ \]
\noindent $ \begin{array}{lll}
 \gamma_{10}^{ns}  &  = & \hspace{3mm} a_s C_F  ( \frac{12055}{1386} )   \\
& &  + a_s^2  \Bigl[
    C_F C_A ( \frac{19524247733}{523908000} )
  + C_F^2  (  - \frac{9579051036701}{1331250228000} )
   +  n_f C_F (  - \frac{2451995507}{288149400} )   \Bigr] \\
& &  + a_s^3  \Bigl[
      C_F C_A^2 ( \frac{94091568579766453}{435681892800000} 
  + \frac{151796299}{8004150} \zeta_3 )
   +  C_F^2 C_A  (  - \frac{16389982059548833}{465937579800000} 
         - \frac{151796299}{2668050} \zeta_3 ) \\
& & \hspace{5mm}
   +  C_F^3 (  - \frac{2207711300808736405687}{127866318149354400000} +
         \frac{151796299}{4002075} \zeta_3 )  
   +  n_f C_F C_A  (  - \frac{9007773127403}{389001690000}
           - \frac{48220}{693} \zeta_3 ) \\
& & \hspace{5mm}
   +   n_f C_F^2  (  - \frac{75522073210471127}{1230075210672000} 
             + \frac{48220}{693} \zeta_3 )
   +   n_f^2 C_F (  - \frac{27995901056887}{11981252052000} )  \Bigr] \\
& = & \hspace{3mm} a_s  11.5969215969 \\
& &+ a_s^2  \Bigl( 136.2741775192
                 - 11.3459453418 n_f \Bigr) \\
& &+ a_s^3  \Bigl(  2379.9199516669 
      -387.6422968015 n_f
      -3.1155231451 n_f^2 \Bigr)  \\
\end{array} $ \\
\vspace{.3cm}

\noindent $ \begin{array}{lll}
     C^{ns}_{2,10} &  = & 1 \\ & &
       + a_s C_F   ( \frac{2006299}{138600} ) \\ & &
       + a_s^2 n_f C_F   (  - \frac{561457267429757}{15975002736000} ) \\ & &
       + a_s^2 C_F C_A   ( \frac{6124093193824187}{29045459520000}
   - \frac{104674}{1155} \zeta_3 ) \\ & &
       + a_s^2 C_F^2   ( \frac{558708799987324013}{14760902528064000}
 + \frac{88798}{1155}
         \zeta_3 ) \\ & &
      + a_s^3 fl_{11} n_f C_F C_A   (  - \frac{3753913187503}{58677696000}
   - \frac{162776}{202125} \zeta_3 + \frac{896}{11} \zeta_5 ) \\ & &
       + a_s^3 fl_{11} n_f C_F^2   ( \frac{3753913187503}{22004136000}
  + \frac{1302208}{606375} \zeta_3 - \frac{7168}{33} \zeta_5 ) \\ & &
       + a_s^3 n_f C_F C_A   (
    - \frac{21664244926039357214987}{23550349033411200000}
  + \frac{10519793104}{42567525} \zeta_3 - \frac{24110}{693} \zeta_4 ) \\ & &
    + a_s^3 n_f C_F^2   (
     - \frac{1521387460036994061010049}{2720065313358993600000}
   - \frac{3997754476}{42567525} \zeta_3 + \frac{24110}{693}
         \zeta_4 ) \\ & &
       + a_s^3 n_f^2 C_F   ( \frac{57084428047851551911}{996360920644320000}
  + \frac{48220}{18711} \zeta_3 ) \\ & &
       + a_s^3 C_F C_A^2   ( 
  \frac{709221119965457939095237}{235503490334112000000} -
       \frac{14713925739913}{6243237000} \zeta_3
  + \frac{151796299}{16008300} \zeta_4 + \frac{190858}{231} \zeta_5 ) \\ & &
       + a_s^3 C_F^2 C_A   (
  \frac{16350009304926933389608829}{8369431733412288000000}
        + \frac{1430215936081}{6163195500} \zeta_3
  - \frac{151796299}{5336100} \zeta_4 - \frac{22658}{99} \zeta_5 ) \\ & &
       + a_s^3 C_F^3   (
   - \frac{3247779532370920623770610131}{92155812816602703168000000} 
  + \frac{2182208825245282}{1622461215375} \zeta_3 +
         \frac{151796299}{8004150} \zeta_4 - \frac{75212}{99} \zeta_5 )  \\ 
& = & 1 + a_s  \Bigl( 19.3006156806 \Bigr)   \\
& & + a_s^2  \Bigl( 639.2106629599 - 46.8613184159 n_f \Bigr)  \\
& & + a_s^3  \Bigl(
    -14.4587445075 n_f fl_{11}
    +24953.1349702005         \\ & &
    -3770.1021201303 n_f
    +80.5209797251 n_f^2
 \Bigr) 
\end{array} $
\vspace{.3cm}
 
\noindent $ \begin{array}{lll} 
  C^{ns}_{L,10}  & = & \hspace{3mm}
        a_s C_F   ( \frac{4}{11} ) \\ & &
  + a_s^2 n_f C_F   (  - \frac{163679}{114345} ) \\ & &
  + a_s^2 C_F C_A   ( \frac{89670761}{8731800} - \frac{48}{11} \zeta_3 ) \\ & &
  + a_s^2 C_F^2   (  - \frac{1999510607}{528273900} 
                      + \frac{96}{11} \zeta_3 ) \\ & &
  + a_s^3 fl_{11} n_f C_F C_A   ( \frac{5073093424963}{88016544000} 
                            + \frac{3641546}{121275} \zeta_3
                             - \frac{960}{11} \zeta_5 ) \\ & &
       + a_s^3 fl_{11} n_f C_F^2   (  - \frac{5073093424963}{33006204000}
           - \frac{29132368}{363825} \zeta_3 + \frac{2560}{11} \zeta_5 ) \\ & &
       + a_s^3 n_f C_F C_A   (  - \frac{176183576988227323}{1699159381920000} +
         \frac{55485434}{1216215} \zeta_3 ) \\ & &
       + a_s^3 n_f C_F^2   ( \frac{9048874326307637}{190368782604000}
                             - \frac{1174256}{15015} \zeta_3 ) \\ & &
       + a_s^3 n_f^2 C_F   ( \frac{63272639}{11320155} ) \\ & &
       + a_s^3 C_F C_A^2   ( \frac{2366034921481985137}{6796637527680000} -
      \frac{95022195887}{187297110} \zeta_3 + \frac{3760}{11} \zeta_5 ) \\ & &
    + a_s^3 C_F^2 C_A   (  - \frac{323139848004267269}{3354750574560000}
    + \frac{22904191}{17325} \zeta_3 - \frac{14240}{11} \zeta_5 ) \\ & &
       + a_s^3 C_F^3   (  - \frac{887562386698645967383}{3166213592269728000} -
       \frac{357031607224}{468242775} \zeta_3 + \frac{13440}{11} \zeta_5 )  \\
& = &  a_s  \Bigl(  0.4848484848 \Bigr)   \\
& & + a_s^2  \Bigl(  32.0176594698 - 1.9085982480 n_f \Bigr)  \\
& & + a_s^3  \Bigl( -2.3976416945 n_f fl_{11}
     +2081.2132221274  \\ & &
      -278.0172176870 n_f  
     +7.4525056120 n_f^2 \Bigr) 
\end{array} $ \\

\section{Acknowledgements}
We are grateful to W.L. van Neerven and J. Smith for helpful discussions.
S.A. Larin gratefully acknowledges the support of the NWO
project ``Computer Algebra and Subatomic Physics''
and of the International Science Foundation, Grants No. N6M000,
N6M300. We also want to thank the CAN foundation for the use of their
computers.

\end{document}